\documentclass[11pt]{article}
\usepackage{amsmath,amssymb,amsthm}
\usepackage[margin=1.3in]{geometry}
\usepackage[round]{natbib}
\usepackage[most]{tcolorbox}
\usepackage{xcolor}
\usepackage{authblk}
\usepackage{tocloft}
\definecolor{darkblue}{rgb}{0.00,0,0.6}
\usepackage[colorlinks=true, citecolor=darkblue,linktocpage=true]{hyperref}
\usepackage{xspace}

\newcommand{\E}{\mathbb{E}}

\newcommand{\calF}{\mathcal{F}}

\newcommand{\calH}{\mathcal{H}}
\newcommand{\cH}{\mathcal{H}}
\newcommand{\cE}{\mathcal{E}}
\newcommand{\commentout}[1]{}
\renewcommand{\P}{\mathbb{P}}
\newcommand{\BF}{\textnormal{BF}}
\renewcommand{\d}{\textnormal{d}}
\newcommand{\glr}{\textnormal{GLR}}
\newcommand{\lr}{\textnormal{LR}}

\theoremstyle{definition}
\newtheorem{example}{Example}
\definecolor{excolor}{RGB}{250, 249, 235}
\tcolorboxenvironment{example}{
  enhanced,
  breakable,
  colback=excolor,
  colframe=white,
  boxrule=0pt,
  arc=0pt,
  left=6pt,
  right=6pt,
  top=6pt,
  bottom=6pt,
  before upper={\setlength{\parindent}{15pt}}
}

\newif\ifshowrevisions
\showrevisionsfalse
\definecolor{revcolor}{rgb}{0.00,0.20,0.85}
\newcommand{\rev}[1]{\ifshowrevisions{\color{revcolor}#1}\else#1\fi}
\newcommand{\revB}[1]{\ifshowrevisions{\color{red}#1}\else#1\fi}
\title{E-values as statistical evidence: A comparison \\ to Bayes factors, likelihoods, and p-values\thanks{Accepted for publication in \emph{Synthese}.}}
\author[1]{Ben Chugg}
\author[1]{Aaditya Ramdas}
\author[2]{Peter Gr\"unwald}
\affil[1]{\small Carnegie Mellon University}
\affil[2]{Leiden University and Centrum Wiskunde \& Informatica}
\date{\today}

\begin{document}

\maketitle

\begin{abstract}
A recurring debate in the philosophy of statistics concerns what, exactly, should count as a measure of evidence for or against a given hypothesis. P-values, likelihood ratios, and Bayes factors all have their defenders. In this paper we add two additional candidates to this list: the e-value and its sequential analogue, the e-process. E-values enjoy several desirable properties as measures of evidence: they combine naturally across studies, handle composite hypotheses, provide long-run error rates, and admit a useful interpretation as the wealth accrued by a bettor in a game against the null distribution. E-processes additionally handle optional stopping and optional continuation. This work examines the extent to which e-values and e-processes satisfy the evidential desiderata of different statistical traditions,
concluding that they combine attractive features of p-values, likelihood ratios, and Bayes factors, and merit serious consideration as interpretable and intuitive measures of statistical evidence.
\end{abstract}

% \begin{center}
% \textbf{Keywords:} e-values, evidence, betting, Bayes factors, likelihood ratios, p-values
% \end{center}

{
\small
\tableofcontents
}

\section{Introduction}

It is sometimes asserted that the p-value, first introduced by Karl Pearson~\citep{pearson1900on} but popularized and developed by Ronald Fisher~\citep{fisher1925statistical,fisher1935design}, is a measure of evidence against the null hypothesis. There is---to put it mildly---significant disagreement over this claim~\citep{berger1987testing,wright1992adjusted,wagenmakers2007practical, hubbard2008p,greenland2019valid,muff2022rewriting,lakens2022p}.

P-values lack several properties that many would expect of a measure of evidence, including consistency, sample size invariance, and the ability to handle optional stopping.
These drawbacks have caused many statisticians to renounce p-values as appropriate measures  of statistical evidence (though, of course, they remain useful for other purposes).  For example, Principle 6 in the American Statistical Association's statement on p-values reads ``by itself, the p-value does not provide a good measure of evidence regarding a model or hypothesis''~\citep{wasserstein2016asa}.

Critics of p-values turn to other proposed notions of evidence, often the likelihood ratio and the Bayes factor. No single notion has satisfied everyone, however, and debates about the best object for quantifying statistical evidence continue to rage (e.g., \citealt{royall1997statistical,forster2006counterexamples,lele2004evidence,taper2016evidential,mayo2011error}).

In this article we do not attempt to resolve these disputes. Instead, we hope to simply add two new candidates to the discussion about statistical evidence: \emph{e-values/e-variables} and \emph{e-processes}.
These objects (together, \emph{e-statistics}) are  simple yet rich objects in mathematical statistics that have garnered significant attention over the past five years. They have proven indispensable to recent progress across several areas, such as multiple testing, mean estimation, and changepoint detection.
We refer to \citet{grunwald2024safe} and \citet{ramdas2025hypothesis} for more on such applications.

\paragraph{\textnormal{\emph{Contributions and outline.}}}
We investigate the extent to which e-statistics satisfy various evidential criteria commonly used in the philosophy of statistics. To do so,
Section~\ref{sec:desiderata} collects 21
desiderata that a measure of statistical evidence should satisfy according to several schools of thought, including Bayesian confirmation theory and various flavors of frequentism.
Section~\ref{sec:comparison} then evaluates e-statistics against these criteria and compares them with p-values, likelihood ratios, and Bayes factors.

While e-statistics do not satisfy all criteria for evidence that we identify throughout the literature, we find that they perform admirably well across a variety of measures.  This stems partly from the fact that they can be viewed as generalized likelihood ratios, thereby inheriting many of the likelihood ratio's evidential properties. But they also satisfy criteria that the likelihood ratio fails, such as providing meaningful evidence
against a single hypothesis without necessarily requiring an explicit alternative hypothesis.\footnote{Though whether this is valuable, or even sensible, depends partly on one's view of statistical evidence.}
That said, the flexibility of e-values and e-processes also has drawbacks. We discuss these in Section~\ref{sec:counterarguments}, where we also discuss their relationship to Birnbaum's theorem and the likelihood principle~\citep{birnbaum1962foundations}.

Beyond relating e-statistics to the literature on statistical evidence, a secondary contribution of the paper is to clarify and sharpen several of the desiderata themselves. In particular, we distinguish between \emph{static} and \emph{dynamic} evidential criteria. Static criteria are evaluated on a fixed batch of data, whereas dynamic criteria concern sequential and counterfactual settings in which additional data may be collected or further studies run. We further distinguish between different kinds of counterfactual scenarios and different forms of optional continuation, providing examples of each. We hope this taxonomy clarifies a set of issues that are often conflated in the literature.

We begin with an overview of e-statistics.

\section{E-Statistics and the Betting Game}
\label{sec:e-statistics}

An e-variable $E$ for a statistical hypothesis (i.e., collection of distributions) $\calH$ is a nonnegative random variable whose expectation is at most 1 under all $P\in \calH$:
\begin{equation}
 \E_\calH[E] := \sup_{P\in\calH}\E_P[E]\leq 1.
\end{equation}
We sometimes differentiate between the random variable $E$, which we call the e-variable, and its realization which we call the e-value. In practice, there is a random quantity $X$ which represents our data (for example $X$ may be a vector of 100 independent observations in an experiment, or a summary of these data such as a $z$-score), and $E$ can be written as a (measurable) function of $X$, i.e. $E$ is a {\em statistic}.

An e-process is the sequential analogue of the e-value. Here, we model our data as a sequence $X_1, X_2, \ldots$ (which we abbreviate henceforth to $(X_t)_{t\geq 1}$) where each $X_i$  may be either a simple data point or a vector. Now consider a sequence of random variables $(E_t)_{t\geq 1} = E_1, E_2, \ldots$
where each $E_i$ can be written as a function of the first $i$ outcomes, $X_1, \ldots, X_i$. We call such a sequence
 an e-process for $\calH$ if  all $E_i$ are nonnegative and $\E_\calH[E_\tau]\leq 1$ for every stopping time $\tau$. Informally, a stopping time is a random (that is, data-dependent) time determined by a stopping {\em rule\/} that for sequences $X_1, \ldots X_t$ of arbitrary length either says  `stop' or `continue'. \rev{Crucially, whether we stop at time $t$ must be decided on the basis of $X_1,\dots,X_t$ alone, and not on data yet to arrive. The realized stopping time $\tau$ is then the first $t$ at which the rule says `stop', and is random because it is a function of the random data.} Examples are `stop at $t=5$', `stop as soon as you have seen two outcomes larger than 1', or `stop at the smallest $t$ such that the p-value based on $t$ outcomes is smaller than $0.05$.'\footnote{Formally, what we covered here is just a special case of the more general, measure-theoretic definition of e-processes and stopping times. There we have a set of filtered probability spaces $\{(\Omega, (\calF_t)_{t\in\mathbb{N}},P)\}_{P\in\calH}$ on which $(E_t)$ is an adapted process (i.e., $E_t$ is $\calF_t$-measurable for each $t$). Recall that a stopping time $\tau$ is defined relative to such a filtration as a random variable such that $\{\tau = t\}$ is $\calF_t$-measurable for each $t$.}

\rev{It is worth flagging immediately that a sequence of e-variables is not in general an e-process. That is, validity at each fixed time does not imply validity at data-dependent times, since one could always stop at whichever $E_t$ happens to be largest. What \emph{is} guaranteed is that a running product of e-variables, each of which is valid conditionally on the past, forms an e-process. We make this precise in Example~\ref{ex:research_group} and the discussion following it, and return to the gap between the two notions in Section~\ref{sec:dynamic-criteria}.}

E-statistics have long appeared implicitly in the statistics and probability literature, especially in relation to nonnegative supermartingales (which are a special case of e-processes). It is only recently however (roughly 2020; see \citealt[Section 1.8]{ramdas2025hypothesis}) that they have begun to be studied as objects in their own right.

E-statistics have often been referred to as measures of evidence against $\calH$~\citep{shafer2021testing,ramdas2023game}: the larger the value the more evidence. This view is buoyed by their interpretations as betting scores in a game between the statistician and nature~\citep{shafer2005probability,shafer2019game}. This game proceeds as follows.

\subsection{The betting game}
Fix some distribution $P \in \calH$. At time $t=1,2,\dots$, the statistician designs a nonnegative random variable $B_t$, which is required to be a function of data $X_1, \ldots, X_t$ arriving at or before time $t$, satisfying\footnote{More formally and generally we can write $\E_P[B_t | \calF_{t-1}]\leq 1$ where $\calF_{t-1}$ is all the information available until and including time $t-1$, which is typically (though not always) the $\sigma$-algebra generated by data $X_1,\dots,X_{t-1}$. \revB{For example, the process $B_t$ in Example~\ref{ex:ttest} depends on a coarser filtration. Strictly speaking $X_i$ in (\ref{eq:betting-score}) should be be replaced by $B_i$. The subtle difference, elaborated on by \citep{perez2024statistics}, is beyond the scope of this paper.} 
}
\begin{equation}
\label{eq:betting-score}
    \E_P[B_t | X_1,X_2,\dots,X_{t-1}]\leq 1.
\end{equation}
We imagine the statistician buying a share of $B_t$ for \$1 and his payoff is $\$ B_t$ once its value is revealed. The inequality in \eqref{eq:betting-score} expresses that, under $P$, the statistician does not expect to make any money, regardless of how cleverly $B_t$ is designed.
We assume that the statistician can buy an arbitrary (nonnegative) number of shares. If the statistician starts with \$1 in round 1, then after the first outcome $X_1$ is revealed he will have \revB{$\$ B_1$}. If he reinvests all this money in round 2, then after $X_2$ is revealed he will have $B_1 B_2$. Thus, if the statistician starts with \$1 and re-invests his money at each point in time, his total capital at time $t$ is given by\footnote{In the e-process literature, $K_t$ is called a
\emph{test supermartingale}.}
$K_t := \prod_{i\leq t} B_i$.

The intuition behind the betting game is that large values of $K_t$ may be considered evidence against $P$. To see this, note that if  data were truly drawn according to $P$, then the statistician does not expect to gain any money in this game, regardless of whatever stopping time $\tau$ is employed. More formally, for any stopping time $\tau$:
\begin{equation}\label{eq:stoppingtime}
    \E_P[K_{\tau}]\leq 1,
\end{equation}
which follows from~\eqref{eq:betting-score} and the optional stopping theorem.\footnote{\rev{The optional stopping theorem makes precise the idea that a fair game cannot be made favorable by a clever choice of when to quit. Condition~\eqref{eq:betting-score} says that $\E_P[K_t \mid X_1,\dots,X_{t-1}] = K_{t-1}\E_P[B_t \mid X_1,\dots,X_{t-1}] \leq K_{t-1}$, i.e.\ that $(K_t)_{t\geq 0}$ is a nonnegative \emph{supermartingale}, so in expectation the capital never grows. The theorem states that for such a process, $\E_P[K_\tau]\leq \E_P[K_0]$ for every stopping time $\tau$. Since the statistician starts with one dollar, $K_0=1$, this gives~\eqref{eq:stoppingtime}. The nonnegativity of $K_t$ is what allows $\tau$ to be unbounded.}} Next, note that it is unlikely that $K_t$ can become large.
How unlikely, exactly? By Markov's inequality, (\ref{eq:stoppingtime}) immediately gives that $P(K_{\tau} \geq 1/\alpha)\leq \alpha$.  Further, by Ville's inequality~\citep{ville1939etude}, a time-uniform extension of Markov's inequality,
\begin{equation}\label{eq:ville}
P\left( \text{there exists\ } t\geq 1: K_t \geq \frac{1}{\alpha} \right)\leq \alpha.
\end{equation}
That is, the probability of the process $(K_t)_{t\geq 1}$ ever exceeding $1/\alpha$, \emph{no matter how long we continue the game}, is at most $\alpha$.

We can relate e-variables and e-processes to the betting game as follows.
When $\calH=\{P\}$ is a singleton (which we henceforth call a \emph{point} or \emph{simple} hypothesis),
an e-variable may be viewed as the payoff from a single-round game, $E = K_1$. The data $X_1$ in this game need not be a single observation---it may itself be a batch of data presented all at once. An e-process, meanwhile,  corresponds to an evolving capital process $(K_t)_{t\geq 1}$.

For composite $\calH$, for each $P \in \calH$ let $(K_t^P)_{t \geq 1}$ be the capital process of a game played against the singleton distribution $P$. Now every e-variable corresponds to the \emph{minimum} payoff in a single-round game, $E \leq  \inf_{P\in\calH} K_1^P$. Analogously, it can be shown that every e-process satisfies
$E_t\leq \inf_{P\in\calH} K_t^P$ (and \emph{admissible} e-processes satisfy this with equality; see \citealp{ramdas2022testing}). In other words, an e-process for $\calH$ reports the minimum wealth across many simultaneous betting games, one for each $P\in\calH$, all played using the same data $X_1,X_2,\dots$.

To relate e-statistics to scientific practice, consider the following scenario.

\begin{example}
\label{ex:research_group}
Suppose a research group analyzes an initial batch of data $X_1$---for example, from a clinical trial---and reports an e-variable $S_1 = S_1(X_1)$. Next, perhaps because the initial findings appear promising, the same or another group analyzes a new, independent batch of data $X_2$, reporting an e-variable $S_2$. The process may then continue, with further groups analyzing independent batches $X_3, X_4, \ldots$ and reporting e-variables $S_3, S_4, \ldots$. To combine the accumulating evidence---for instance, in a meta-analysis \citep{SchureG22}---the researchers multiply the successive e-variables.

We may think of the e-variable $S_t$ as the payoff $B_t$ at time $t$. Since the design of the e-variable $S_t = B_t$ at time $t$ may depend on the earlier data $X_1, \ldots, X_{t-1}$, viewing the batches as arriving sequentially turns the defining e-variable condition $\E_P[S_t] \leq 1$ into the conditional requirement $
\E_P[S_t \mid X_1, \ldots, X_{t-1}] \leq 1$.
It follows that, for any stopping time $\tau$, $
E_\tau := \prod_{i \leq \tau} S_i$
is an e-variable, and that the running product $
E_t := \prod_{i \leq t} S_i$ defines an e-process.
\end{example}

Example~\ref{ex:research_group} constructs an e-process by multiplying e-variables. Conversely, when $\calH$ is simple, every 
\revB{admissible}
e-process $(E_t)_{t\geq 0}$ for $\calH$ can be decomposed as a product of `past-conditional' e-variables, i.e. of the form $B_t$ satisfying (\ref{eq:betting-score}).
Explicitly distinguishing between these two construction-directions (creating e-processes out of e-variables and vice-versa) is not always important, yet it will become highly useful in Section~\ref{sec:dynamic-criteria}.

Next let us provide an example of a  fundamental e-statistic.
\begin{example}[Likelihood ratio]
\label{ex:lr}
Consider a simple  hypothesis $\calH=\{P\}$ where $P$ has density $p$. Then the likelihood ratio $q/p$ for any distribution $Q$ with density $q$ is an e-variable for $\calH$ (see below for assumptions implicit in this notation). To see this, note that
    \begin{equation}
 \label{eq:lr-is-eval}
 \E_{P}\left[\frac{q(X)}{p(X)}\right] = \int \frac{q(x)}{p(x)} p(x) \d x = \int q(x)\d x = 1.
 \end{equation}
 By similar reasoning, for a sequence of random variables $(X_t)_{t\geq 1}$,  the process $(L_t)_{t\geq 1}$ where $L_t = q(X_1,\ldots, X_t) /p(X_1,\ldots, X_t)$ defines an e-process for the hypothesis that the data are distributed according to $P$. If further they are independent and identically distributed (iid) under $P$ and $Q$, we can write $L_t = \prod_{i \leq t} q(X_i)/p(X_i)$.
\end{example}

By recovering the likelihood ratio for point hypotheses, Example~\ref{ex:lr} highlights that one useful way to understand e-statistics is as extensions of likelihood ratios to composite settings. In particular, as opposed to some  other popular extensions to the composite setting such as the generalized likelihood ratio (Section~\ref{sec:desiderata}), they choose to preserve the property $\E_\calH[E]\leq 1$, which we can view as ``fairness'' in the betting game: evidence is difficult to exaggerate when playing against the null.
We will return to the connection between \rev{simple-versus-simple} likelihood ratios and general e-statistics later on.

We provide additional examples in Section~\ref{sec:examples}, but to orient unfamiliar readers, let us provide a second e-statistic here. This example can be viewed as an instance of the former.
\begin{example}[Gaussian e-variable]
\label{ex:subgaussian}
For any $\lambda\in\mathbb{R}$, the object $E_\lambda(X) = \exp(\lambda X - \lambda^2\sigma^2/2)$ is an e-variable for $\calH = \{N(0,\sigma^2)\}$, i.e., a Gaussian with mean 0 and variance $\sigma^2$.
Note that $E_\lambda$ is the likelihood ratio between $N(\sigma^2\lambda,\sigma^2)$ and $N(0,\sigma^2)$.
Further, the process $(E_t)_{t\geq 1}$ where $E_t = \exp(\lambda \sum_{i\leq t}  X_i - t \lambda^2\sigma^2/2)=\prod_{i\leq t} \exp(\lambda X_i - \lambda^2\sigma^2/2)$ is an e-process for the hypothesis that the data $(X_t)_{t\geq 1}$ are iid from $N(0,\sigma^2)$.
\end{example}

\noindent\textbf{Remark on notation and definitions:}
Although the theory of e-statistics can be developed in  much greater generality \citep{larsson2025numeraire}, throughout this paper we assume that all distributions admit densities or probability mass functions. We use capital letters such as $P$ and $Q$ to denote distributions, and corresponding lowercase letters such as $p$ and $q$ to denote their densities/mass functions. Whenever we refer to likelihood ratio $q/p$ with $P \in \calH$, we assume  it is almost surely well-defined under both $\calH$ and $Q$.

\subsection{Evidence for or against?}
\paragraph{The meaning of small $E$.} The betting game provides a notion of evidence \emph{against} a hypothesis $\calH$, as opposed to a notion of evidence \emph{for} the hypothesis. In particular, if $E$ is an e-variable for $\calH$, then in general small values of $E$ need not provide evidence in favor of $\calH$. This is because one can play the betting game conservatively. As an extreme example, the bet $B_t\equiv 1$ is an e-variable but will never make any money regardless of the discrepancy between $\calH$ and the true data generation process.

That said, e-statistics are often used to compare two hypotheses. In hypothesis testing, for instance, we compare a null $\calH_0$ to an alternative $\calH_1$ and search for e-statistics under $\calH_0$ which grow quickly under $\calH_1$. For example, for singletons $\calH_0 = \{P_0\}$ and $\calH_1 = \{P_1\}$, we can consider the likelihood ratio (Example~\ref{ex:lr}).  This suggests that there are indeed cases when a small e-statistic $E$ for $\calH_0$ should be viewed as evidence in favor of $\calH_0$.

\rev{A necessary requirement for such a scenario is that the reciprocal $E^{-1}$ of $E$ be an e-statistic when we reverse the roles of $\calH_0$ and $\calH_1$, as is the case for the likelihood ratio. It is not, however, sufficient. Example~\ref{ex:ryabko} below exhibits an $E$ whose reciprocal is a perfectly good e-statistic for the reversed problem, and yet whose alternative is not one that anybody should endorse. When the condition holds \emph{and} the alternative is a bona fide hypothesis about the world, evidence against $\calH_0$ can indeed be viewed as evidence for $\calH_1$ and vice versa.}

\paragraph{The meaning of large $E$.}
Often, however, $\calH_1$ is implicit or chosen pragmatically based on the problem at hand. This is the case in recent work on bounded mean testing \citep{waudby2024estimating} (Example~\ref{ex:bounded-means} below) or in various conformal e-statistics \citep{vovk2025conformal}, for instance.
In such cases, the perspective of a large e-value providing evidence against $\calH_0$ remains clear whereas the interpretation as evidence for a specific $\calH_1$ less so. The following example shows that this may happen even when $\calH_0= \{P\}$ is simple and the e-statistic $E$ is a likelihood ratio
\revB{.} 
%so that formally, $E^{-1}$ is an e-statistic as well.
In general, while an alternative can be inferred from the bets or, more indirectly, from the e-statistic's definition, such an alternative might be best viewed as instrumentally useful, and not representing a bona fide hypothesis that represents a true state of the world.
\begin{example}\label{ex:ryabko}
\cite{ryabko2005using} show that bit strings produced by standard
random number generators can be substantially compressed by
lossless data compression algorithms such as \texttt{zip}, which is a clear
indication that the bits are not so random after all. Here, the null hypothesis states that data are `random' (independent fair coin flips), and one measures `amount of evidence against $\cH_0$ provided by data $x^\tau = x_1, \ldots, x_{\tau}$' as
$$
\tau- \textsc{cl}_{\texttt{zip}}(x^{\tau}),
$$
where $\textsc{cl}_{\texttt{zip}}(x^{\tau})$ is the number of bits needed to code $x^{\tau}$ using (say) \texttt{zip}. Kraft's inequality \citep{CoverT91}  says that for arbitrary \rev{prefix} codes for encoding binary strings of length $\tau$ into other binary strings of non-fixed length, the code length $\textsc{cl}(x^{\tau}) $ as measured in bits satisfies $\sum_{x^{\tau} \in {\cal X}^{(\tau)}} 2^{-\textsc{cl}(x^{\tau})} \leq 1$, where ${\cal X}^{(\tau)}$ is the set of binary strings determined by stopping time $\tau$ (if $\tau=n$ for fixed $n$, this is simply $\{0,1\}^n$).
Thus, if we set $q(x^{\tau}) := 2^{-\textsc{cl}_{\texttt{zip}}(x^{\tau})}$ we get $\sum_{x^{\tau} \in {\cal X}^{(\tau)}} q(x^{\tau}) \leq 1$: $q$ represents a {\em sub-probability distribution}.
At the same time, for the null we have  $\cH_0 = \{P \}$, where $P$ represents a sequence of fair coin flips, so it has mass function $p$ with for each $x^{\tau} \in {\cal X}^{(\tau)}$,  $p(x^{\tau}) = 2^{-{\tau}}$. Defining $E_{\tau}= q(X_1,\ldots, X_{\tau} )/\revB{p(X_1, \ldots, X_{\tau})}$ we thus find
$$
{\mathbb E}_{P}[E_{\tau}] = \sum_{x^{\tau} \in {\cal X}^{(\tau)}} \frac{q(x^{\tau})}{p(x^{\tau})} p(x^{\tau}) \leq 1\text{~ and ~}
\log E_{\tau} = \tau - \textsc{cl}_{\texttt{zip}}(X_1, \ldots, X_{\tau}).
$$
Thus, the Ryabko-Monarev bit-difference is the logarithm of
an e-value. But note that there is no explicitly defined alternative. Being able to significantly compress a string by \texttt{zip} intuitively provides strong evidence that the null hypothesis is false, and this is formalized by e-processes. Nevertheless, even though we measure evidence by the log likelihood ratio between $\calH_0=\{P\}$ and $Q$, this evidence against $\{P\}$ should clearly  {\em not\/} be construed as  evidence in favor of the sub-distribution $Q$, which is a complicated object that helps to {\em detect\/} some structure in $x_1, \ldots, x_{\tau}$ but is not best thought of as either generating or predicting that structure.
\end{example}
It turns out that this example generalizes: any e-statistic has a code length-difference interpretation, which connects it to notions of evidence that have implicitly been suggested by the information theory community, often in the context of {\em Minimum Description Length\/} model comparison \citep{GrunwaldR20,Grunwald07}. Since it is less well-known, we do not spell out the desiderata coming from this tradition here but refer instead to \citep{GrunwaldR20} for those interested.

To summarize: e-statistics are best regarded as measures of evidence against a designated hypothesis, with ``evidence for an alternative'' emerging only in distinctive scenarios.

\subsection{Optimality of e-statistics}
Example~\ref{ex:ryabko} notwithstanding, in some settings there is a clearly defined alternative hypothesis $\calH_1$. In such cases, it is natural to call an e-variable for testing $\calH_0$ {\em optimal} if it tends to become large under $\calH_1$. In particular, if the data are in fact generated under $\calH_1$, we would like evidence against $\calH_0$ to accumulate quickly. This idea of ``growing quickly under the alternative'' can be formalized in several ways. The most natural, and by far the most studied, is {\em log-optimality}; the e-variable achieving it has been called {\em GRO} (growth-rate optimal) \citep{grunwald2024safe,lardy2024reverse} or the {\em numeraire} \citep{larsson2025numeraire} (the latter paper also describes alternative notions of optimality, as does \citealt{koning2024continuous}). In fact, the likelihood ratio from Example~\ref{ex:lr} is the log-optimal e-variable for testing ${P}$ against ${Q}$. More broadly, this highlights that a given hypothesis may admit many different e-variables and e-processes. We return to this point in Section~\ref{sec:counterarguments}. For now, let us turn to a discussion of the various statistical traditions and what properties they believe a measure of evidence should possess.

\section{An Overview of Evidential Desiderata}
\label{sec:desiderata}

Unsurprisingly given the fierce debates at the foundations of statistics, there isn't a single checklist of evidential criteria adopted by all statisticians and philosophers of statistics alike.  Instead, different statistical traditions argue for different criteria. Following \citet{forster2006counterexamples}, we group these traditions into three broad groups:
\begin{enumerate}
    \item[(i)] likelihood-based accounts, which treat evidence as relative support between hypotheses and elevate likelihood ratios as fundamental;
    \item[(ii)] Bayesian accounts, which quantify evidence by changes in model odds, typically via Bayes factors and their approximations; and
    \item[(iii)] error-probability accounts, which connect evidential strength to the ability of procedures to detect discrepancies (e.g., through type-I/II error rates or confidence intervals).
\end{enumerate}
The gray box below summarizes the main criteria proposed by these three constellations of statistical philosophies, augmented with some of our own. But let us first provide some more background on these three traditions. We begin with the latter.

\paragraph{The error-probability tradition.}
The error-probability tradition traces back to Fisher, Neyman, and Pearson.
There is no unified view within this tradition of how statistics should be practiced. In fact, Fisher famously had significant disagreements with Neyman and Pearson, who proposed a fully decision-theoretic theory of statistics in which `evidence' really plays no role---there are only decisions, risks and error probabilities. Nevertheless,  Fisher, Neyman and Pearson were all frequentists
and informal references to `evidence' within the Neyman-Pearson tradition abound.\footnote{In fact, even \cite{Neyman76} himself wrote `my own preferred substitute for `do not reject $H$' is `no evidence against $H$ is found'.}
And, despite these disagreements, contemporary frequentist practice largely draws on a hybrid error-probability tradition combining ideas from Fisher, Neyman, and Pearson.
We can thus still identify core assumptions on the desired properties within this tradition of a measure of evidence. In particular, a measure of evidence should bound or diagnose the probability of being misled.

In other words, the evidential strength of a given object or procedure is inseparable from its (frequentist) control of various error rates (type I/II error, risk control, false discovery control, etc.) Evidence is stronger when the method would rarely give such supportive results if the underlying claim were false. This is precisely the guarantee offered by a p-value, which
is a statistic $T=T(X)$ such that $P(T\leq \alpha)\leq \alpha$ for all $\alpha\in(0,1)$ and all $P\in\calH$. (Here we would say that $T$ is a p-value for $\calH$.)\footnote{\rev{Readers more familiar with the textbook formulation---``the probability of obtaining test results at least as extreme as the result actually observed, assuming the null hypothesis is correct''---may find it helpful to see how the two are related. Suppose evidence against $\calH$ is summarized by some test statistic $S(X)$, with larger values counting as more extreme, and define the tail probability $T(x) := \sup_{P\in\calH} P\big(S(X)\geq S(x)\big)$. This $T$ is precisely the textbook p-value, and a short calculation shows that it satisfies $P(T\leq\alpha)\leq\alpha$ for every $\alpha$ and every $P\in\calH$. The definition above simply isolates that property---which is what all the error-probability guarantees actually rest on---and drops the requirement that $T$ arise from any particular $S$.}} Thus, on the evidential view of a p-value, small values of \rev{$T$} are  evidence against $\calH$.\footnote{\rev{We say ``on the evidential view'' advisedly. Many within the error-probability tradition would deny that the p-value is a measure of evidence at all---for the strict Neyman-Pearsonian it is an input to a decision rule and nothing more---and would therefore regard the question we pose in this paper as not quite the right one to ask. However, our aim is not to adjudicate whether each tradition ought to be in the business of measuring evidence, but to take the criteria each has articulated and ask how e-statistics fare against them. Readers who reject the evidential reading of p-values may simply read the corresponding rows of Table~\ref{tab:comparison} as reporting what the p-value would deliver were it forced into evidential service.}}

Extensions of p-values such as s-values \citep{greenland2019valid}, second-generation p-values \citep{blume2019introduction}, and replication values \citep{killeen2005alternative} are all rooted in the error-probability tradition. Confidence curves/distributions are also sometimes proposed as evidential summaries on this view~\citep{xie2013confidence}. The error-statistics framework of Mayo and Spanos~\citep{mayo2011error,mayo1996error} (which downplays the differences between Fisher and Neyman and focuses on commonalities instead) and Allan Birnbaum’s ``confidence concept of evidence'' \citep{birnbaum1977neyman} likewise sit squarely in the error-probability family.

\paragraph{Bayesian confirmation theory.} The Bayesian statistical tradition is, naturally, rooted in the Bayesian view of probability, which treats model parameters as random variables instead of unknown constants. An agent places an initial prior distribution $\pi$ over parameters $\theta$ (often referred to as the agents' \emph{beliefs}) and, upon observing data $X$, updates
$\pi(\theta)$ via Bayes' rule to a posterior $\pi(\theta \mid X)$. Instead of controlling error, evidence on the Bayesian view is linked to
\emph{confirmation}: data provide evidence for a hypothesis or model to the extent that they increase its posterior support relative to its prior support.

The most common tool for quantifying evidence on the Bayesian view is the Bayes factor \citep{jeffreys1939theory,kass1995bayes}, which is typically used to compare two distinct hypotheses. In particular, for $\calH_0=\{P_\theta: \theta\in\Theta_0\}$ and $\calH_1=\{P_\theta: \theta\in\Theta_1\}$ equipped with priors $\pi_0$ and $\pi_1$, we can write the change in posterior odds after observing $X$ as the prior odds multiplied by the Bayes factor:
\begin{equation}
\frac{\P(\calH_1 \mid X)}{\P(\calH_0 \mid X)}
=
\frac{\P(\calH_1)}{\P(\calH_0)} \cdot \text{BF}(X),\text{~ where ~}\BF(X) = \frac{\P(X|\calH_1)}{\P(X|\calH_0)},
\end{equation}
where $\P(X|\calH_j) = \int_{\Theta_j} P_\theta(X)\pi_j(\theta) \d\theta$ is the marginal likelihood under hypothesis $j$. Note that the posterior odds are defined by both prior distributions over hypotheses, denoted as $\P(\calH_j)$, as well as prior distributions over parameters $\Theta_j$, denoted by $\pi_j$. For the purposes of the Bayes factor, however, only $\pi_0$ and $\pi_1$ are necessary to define.

Bayes factors can be difficult to compute and approximations such as BIC are therefore sometimes offered~\citep{wagenmakers2007practical}. However, for the purposes of an evidential standard, exact Bayes factors are the most popular among Bayesian confirmation theorists, and will thus be our main focus here.
For more detailed discussion of various measures of Bayesian confirmation, see~\citet{fitelson1999plurality,eells2000measuring}.

\paragraph{The likelihood tradition.}
Finally, there is the likelihood camp, which emerged as a rival to both Bayesianism, which it saw as too subjective, and error-probability accounts, which it saw as overly concerned with decision-making instead of evidence~\citep{hacking2016logic,royall1997statistical,edwards1972likelihood}. That said, likelihoodists and Bayesians share a number of commitments,  emphasizing criteria such as combination within and across studies, continuity, consistency, and sharing the view that evidence is fundamentally comparative between two hypotheses~\citep{taper2010nature,taper2016evidential,lele2004evidence}. \rev{But likelihoodists side with the frequentists on one decisive point: they want an objective, prior-free measure of evidence, one fixed by the data and the models under consideration rather than by the analyst's degrees of belief. But this is the only sense in which we call them frequentists here since they do not, in general, accept the further frequentist commitment that evidential strength must be underwritten by long-run error rates. }

Unlike the Bayes factor, however, which marginalizes over the prior, the likelihood ratio is not immediately well-defined for composite hypotheses. In that case, one must appeal to some generalization, such as  the generalized likelihood ratio (GLR)
\begin{equation}
\label{eq:GLR}
\glr(\Theta_0,\Theta_1):=
    \frac{\sup_{\theta_1\in\Theta_1}p_{\theta_1}(X)}{\sup_{\theta_0\in\Theta_0} p_{\theta_0}(X)},
\end{equation}
or others (cf.\  \citealt{bickel2011predictive,bickel2012strength}). Indeed, the Bayes factor is one possible extension to the composite setting, though likelihoodists tend to prefer using objective priors~\citep{jeffreys1939theory} in this case. E-statistics, as we will continue to discuss, are another possible generalization.
As a consequence, it is sometimes natural to refer to e-statistics as ``generalized likelihood ratios,'' and we will do so throughout this paper. However, it is important to keep in mind that they are distinct from~\eqref{eq:GLR}.

Aside from these two, the evidential properties of extensions of the likelihood ratio to composite settings are not always clear, and often do not preserve the comfortable logic of the likelihood ratio: \revB{that the observed data are $a$ times as likely under $\theta_1$ as under $\theta_0$ if $p_{\theta_1}(X) =a p_{\theta_0}(X)$}. In Section~\ref{sec:comparison} we focus on the GLR as the natural composite extension, but we trust that readers can extend the logic to a different generalization if desired.\footnote{We note that in many important instances (i.e. choices of $\calH$), \emph{conditional likelihoods} \citep{royall1997statistical} and \emph{partial likelihoods} such as those underlying the \cite{Cox72} regression model do turn out to be e-statistics \citep{hao2024values}.} \\

With this background established, we turn to the evidential desiderata endorsed by these three traditions. \rev{Of course, it should not be assumed that every entry carries equal weight, even among those traditions which endorse it. The list is thus not a scorecard, and a measure which satisfies more entries is not automatically better. Still, the list gives a useful comparison point.   }

As noted in the introduction, we divide these into two categories: static and dynamic. The static criteria are relatively standard in the literature and require little explanation. The dynamic criteria, by contrast, are more nuanced---and in some cases, entirely novel.
We further divide the dynamic criteria into two sub-categories: inter-experiment and intra-experiment. The former concerns the relationship between multiple experiments, while the latter focuses on the internal dynamics of a single experiment. Because these dynamic criteria are subtle, Section~\ref{sec:dynamic-criteria} devotes significant space to clarifying their meaning, comparing them with related criteria, and illustrating them through specific examples.

% \rev{One caveat before we begin. The list that follows should not be read as a scorecard on which every entry carries equal weight, and a measure that satisfies more of them is not thereby better. The desiderata are indexed to the traditions that endorse them---which is what the ``Trad.''\ column of Table~\ref{tab:comparison} records---and seen against the background of those traditions, they are decidedly not on a par. A Bayesian will be untroubled by a failure of long-run error rates, since the guarantee they express is not one Bayesian inference sets out to provide; an error-probabilist will be equally untroubled by a failure of the likelihood principle, and may regard coherence as beside the point. Which failures count as damaging therefore depends on what one already thinks evidence is for. What we take the list to establish is the more modest claim that e-statistics do well by the lights of several traditions at once, rather than the claim that they win on points.}

\begin{tcolorbox}[
    breakable,
  colback=gray!10,      % light gray background
  colframe=gray!10,     % same as background = no visible border
  sharp corners,         % no rounded edges
  before upper=\setlength{\parskip}{0.6\baselineskip},
]
\small
\begin{center}
    \large\textbf{Evidential Desiderata}
\end{center}

\begin{center}
    \underline{Static criteria}
\end{center}

\textbf{Scalar measure.}
Evidence for or against a hypothesis should be quantified by a single real number. This is a criterion shared by all traditions.

\textbf{Continuous measure.} \revB{Evidence should be graded rather than dichotomous. That is, there is no fixed threshold at which something abruptly becomes evidence}\rev{. Rather, evidence comes in degrees}. This is again endorsed by all three traditions.\footnote{Note this is distinct from \emph{decision-making} in the error-probability tradition, which thresholds p-values in order to either reject or sustain the null hypothesis. But as a measure of evidence, the p-value is meant to be continuous.} See, e.g., \citet{royall1997statistical,lele2004evidence}.

\textbf{Irrelevance of the analyst.}
The evidence should depend on the data only and not on whoever is running the experiment. That is, evidence should be objective. This is endorsed by the likelihoodists and error-probabilists, and explicitly rejected by the Bayesians. See \citet{taper2016evidential,royall1997statistical,bickel2012strength}.

\textbf{Coherence.} A measure of evidence should not assign higher evidence to any implication of a hypothesis than to the hypothesis itself~\citep{gabriel1969simultaneous}.
Formally:
If $\calH\subset\calH'$ then you should have at least as much support for $\calH'$ as you do for $\calH$~\citep{schervish1996p}.
Or, equivalently, for $\calH\subset\calH'$, the evidence against $\calH'$ should not exceed that against $\calH$ \citep{bickel2024david}.
This desideratum arose from multiple testing, and was only later applied to criticize p-values as valid measures of evidence \citep{schervish1996p}.\footnote{This is distinct from the notion of coherence often discussed in a Bayesian context, which is the requirement that a set of probabilities be immune from a Dutch book \citep{berger1983defense}.}

\textbf{Consistency.}
A measure of evidence is consistent if, roughly speaking, it favors the true hypothesis as the sample size increases.
This has been formalized in various ways (cf.\ \citealt{grendar2012p,berger2003approximations}).
Here we will say that, when comparing $\calH_0$ and $\calH_1$, a measure of evidence $M_n$ on $n$ observations is consistent if $M_n$ tends to
$\infty$ in probability under $\calH_1$, and tends to $0$ in probability under $\calH_0$.
Consistency is typically promoted by likelihoodists~\citep{royall1997statistical} and Bayesians~\citep{chib2016bayes}.

\textbf{Sample size invariance.}
A given numerical value of the evidence should correspond to the same strength of evidence regardless of sample size or other aspects of the sampling design (i.e., ``same score, same evidence'').
For instance, an evidence value of $x$ obtained with $n=20$ should be comparable to the same value $x$ obtained with $n=200$. This is supported by Bayesians and likelihoodists; see \citet{wagenmakers2007practical,royall1986effect,goodman1988evidence}.

\textbf{Scale invariance.}
One-to-one transformations of the sample space (e.g., unit changes) should not change the evidence. For instance, measuring distance in meters versus feet should not change the interpretation.
This is supported by likelihoodists and Bayesians~\citep{taper2011evidence}.

\textbf{Reparameterization invariance.}
One-to-one transformations of the model parameters
should not change the evidence. This is supported by likelihoodists \citep{hartigan1967likelihood,berger1983defense,lele2004evidence,taper2011evidence} and (some) Bayesians \citep{jeffreys1946invariant,ghosh2011objective}.

\textbf{Single hypothesis validity.}
One should be able to give evidence for,  or at least
against, a single hypothesis, without comparing it to another hypothesis. This is explicitly endorsed by the error-probability tradition, and explicitly rejected by the likelihood tradition \citep{edwards1972likelihood,royall1997statistical}.

\textbf{Long-run error rates.}
Thresholds on the evidential scale should correspond to explicit bounds on the long-run frequency with which they are exceeded when the hypothesis is true.
This is the main desideratum of the error-probability tradition. See, e.g., \citet{neyman1928use,mayo1996error,mayo2018statistical,mayo2011error}.

\textbf{Composite hypotheses.} A measure of evidence should handle composite hypotheses.
This is implicitly endorsed by Bayesians, who place priors over composite hypotheses, error-probabilists, and by those likelihoodists who seek to generalize the standard likelihood ratio. It has also motivated a significant amount of recent work in the e-statistics literature~\citep{ramdas2023game}.

\textbf{Nonparametric hypotheses.} A measure of evidence should handle nonparametric hypotheses, where likelihood ratios may be very hard to define (due to technical reasons, like not having reference measures to define densities). As above, this is implicitly endorsed by many proponents of most traditions, but some of them handle it significantly more satisfactorily than others.

\textbf{The likelihood principle.}
If two data sets induce proportional likelihood functions for the parameters of interest, then they should carry the same evidence about those parameters.
This principle is foundational for the likelihood tradition \citep{edwards1972likelihood,royall1997statistical,forster2004likelihood,berger1988likelihood} and is also adopted by Bayesians who interpret likelihood function as marginal likelihood.

\textbf{Composability under dependence.}
It should be possible to combine evidence from multiple sources without requiring strong assumptions on their dependence structure. Such a desideratum arises more from practice than from philosophy. In multiple testing, for instance, it is common to want to combine p-values which can rarely be considered independent~\citep{vovk2022admissible,efron2012large}.
But variants of composability are supported by both the likelihoodist and Bayesian traditions. See \citet{royall1997statistical,bickel2012strength,morey2016philosophy}.

\begin{center}
    \underline{Inter-experiment dynamic criteria}
\end{center}

\textbf{Accumulation I: Fixed design.}
When data arise from a fixed number of independent experiments, the total evidence should accumulate from the evidence contributed by the component parts.
This is supported  by both the likelihoodist and Bayesian traditions. See \citet{royall1997statistical,bickel2012strength,morey2016philosophy}.

\textbf{Accumulation II: Flexible design.}
When data arise from a sequence of independent experiments, and the decision to perform the next experiment may depend on the result of the previous one, the total evidence should accumulate from the evidence contributed by the component parts. This generalizes Accumulation I by allowing the number of experiments to be data-dependent. The criterion is discussed frequently in the literature on e-values, where it serves as part of their motivation \citep{grunwald2024safe}.

\textbf{Dynamic consistency.}
When assessing the evidence in an overall sample, we should obtain the same value regardless of whether we analyze the sample as a whole or partition it into sub-samples and combine the results. Variations of this desideratum have been studied extensively in the economics and decision-theory literature~\citep{EpsteinS03,GrunwaldH11}. \rev{Despite the surface similarity, this is a strictly stronger demand than Accumulation I. Accumulation asks only that the pieces \emph{can} be combined into a valid measure of the total evidence. Dynamic consistency asks in addition that the combined value \emph{agree} with what one would have obtained by analyzing the pooled sample directly. A measure can therefore satisfy the former and fail the latter, as products of e-variables do (Section~\ref{sec:dynamic-criteria}).}

\textbf{Inter-experiment counterfactuals.}
A measure of evidence should not depend on data that were not observed. More concretely, it should not depend on decisions about whether or not to continue collecting a new batch of data in counterfactual scenarios in which the data were different from what they actually were. \rev{Where Accumulation II concerns a further experiment that \emph{is} run, and asks that its evidence combine properly with what came before, the present criterion concerns a further experiment that is \emph{not} run, but would have been had the first batch come out differently. }

\begin{center}
    \underline{Intra-experiment dynamic criteria}
\end{center}

\textbf{Intra-experiment optional stopping.}
The reason an experiment was stopped should not affect its evidential value. For example, stopping early because the initial results appear promising, or continuing longer than originally planned, should not change the evidence. This is closely related to, but distinct from, Accumulation II: here the issue is whether to stop or continue a single experiment, rather than whether to run an additional one. Handling optional stopping satisfactorily is a major motivation in modern sequential analysis \citep{ramdas2023game}.

\textbf{Intra-experiment counterfactuals.}
The measure of evidence should depend only on the realized event that the observer learns has occurred, and not on unrealized contingencies in the experimental protocol. Thus, if two descriptions of the experiment yield the same realized event $Y$, they should yield the same evidence. This criterion, together with the previous one, has motivated criticism of p-values~\citep{wagenmakers2007practical,forster2004likelihood}. The general issue of counterfactuals is often brought up by likelihoodists.

\end{tcolorbox}

\section{E-statistics, P-values, Likelihood Ratios, Bayes Factors}
\label{sec:examples}

Before investigating to what extent e-statistics satisfy the desiderata listed above, let us provide more detail on how they relate to p-values, likelihood ratios, and Bayes factors. We begin with p-values.

E-values can be converted to p-values and vice-versa. Indeed, for an e-value $E$, $1/E$ is a p-value by Markov's inequality: $P(1/E<\alpha) = P(E>1/\alpha)\leq \alpha$. The inverse of a p-value is not an e-value in general. E-values tend to be more conservative than p-values, i.e. more extreme data is needed to reach a threshold $1/\alpha$.
P-values can be converted to e-values via \emph{calibrators} \citep{vovk2021values}: Non-negative functions such that $\E_\calH f(P)\leq 1$. Examples of calibrators include $f(p) = \kappa p^{1-\kappa}$ for any $\kappa\in(0,1)$, and $f(p) = \int_0^1 \kappa p^{\kappa-1 }d \kappa$. Unfortunately, the e-values obtained via calibrators are usually not very powerful. \rev{In particular, they grow far more slowly under the alternative than an e-value designed for the problem at hand}, and good (for example, log-optimal) e-values must typically be designed directly.

Next, let us introduce the following general method for constructing e-variables and e-processes when comparing two composite hypotheses.

\begin{example}[Universal inference]
\label{ex:universal-inference}
Consider independent and identically distributed (iid)  observations $X_1,X_2,\dots,X_t$, null $\calH_0$,  and alternative $\calH_1$, both of which may be composite.
\citet{wasserman2020universal} introduce the universal inference estimator
\begin{equation}
  U_t := \prod_{i=1}^t \frac{\hat{q}_i(X_i)}{\hat{p}_t(X_i)},
\end{equation}
where $\hat{q}_i$ is some density in the alternative $\calH_1$ which may be based on the first $i-1$ samples, and $\hat{p}_t$ is the maximum likelihood estimate among all distributions in the null, based on all $t$ observations.\footnote{
Here, for ease of comparison to other e-processes, we consider the case without a holdout dataset. Universal inference is more commonly known for the case where we do have a holdout dataset, we might compute $\hat{q}_i$ on these data to maximize $U_t$, in which case $\hat{q}_1 = \dots = \hat{q}_t$. This is called ``split universal inference.''}
Alternatively, one can also equip $\calH_1$ with a prior $\pi$ (which must be independent of $X_1,\dots,X_t$) and consider
\begin{equation}\label{eq:chatcatch}
    V_t := \int \prod_{i\leq t} \frac{q(X_i)}{\hat{p}_t(X_i)} \d\pi(q),
\end{equation}
which we call the ``method of mixtures.''
To see that $U_t$ is indeed an e-variable for $\calH_0$, note that $\prod_{i\leq t} \hat{p}_t(X_i) \geq \prod_{i\leq t} p(X_i)$ for all densities $p$ in the null by definition, so for all $P\in\calH_0$, $\E_{P} [U_t]\leq \E_P[\prod_{i\leq t} \frac{\hat{q}_i(X_i)}{p(X_i)}] = \int \left(\prod_{i\leq t} \frac{\hat{q}_i(x_i)}{p(x_i)} \right)p(x_1)\cdots p(x_n) \d x_1\cdots \d x_n = \prod_{i\leq t} \int \hat{q}_i(x_i) \d x_i = 1$, using that $\hat{q}_i$ is a probability density.  Similar arithmetic applies for $V_t$. Moreover, $(U_t)_{t\geq 1}$ and $(V_t)_{t\geq 1}$ are e-processes.
\end{example}

When both hypotheses are simple, the universal inference e-variable collapses to the likelihood ratio. While it serves as an illuminating example, we do not advocate its use in all situations. In case of a simple alternative and a composite null, the numeraire e-variable~\citep{larsson2025numeraire} is inherently preferable to the universal inference e-variable (even if we have a different notion of optimality in mind than log-optimality): for all outcomes, the numeraire is at least as large as the universal inference e-variable, whereas for some outcomes the inequality is strict. \rev{The reason this ordering is decisive is that both objects are e-variables for the same null, and so carry exactly the same type-I error guarantee. Universal inference is therefore \emph{inadmissible} in this setting, in the usual decision-theoretic sense of being weakly dominated.} On the other hand, \revB{the numeraire e-variable} defines, in general, an e-variable only and not an e-process.\footnote{By this we mean that if we construct the numeraire e-variable $E_1, E_2, \ldots$ with $E_i$ the numeraire variable for sample $(X_1, \ldots, X_i)$, the resulting process $(E_t)_{t\geq 1}$ is not an e-process \cite{ramdas2023game}.}

The method of mixtures---pioneered by Herbert Robbins in the context of obtaining confidence sequences~\citep{robbins1970statistical}---and the method of predictable plug-ins, respectively illustrated by \rev{$V_t$ and $U_t$} above, are common strategies in the world of e-statistics and illustrate how e-statistics can serve as a bridge between Bayesians and frequentists. Indeed, while e-statistics are fundamentally frequentist objects, the method of mixtures shows how they can accommodate priors while retaining their frequentist properties. Meanwhile, the method of predictable plug-ins can be seen as learning a distribution over the alternative over time, which also has a very Bayesian spirit\rev{: the plug-in $\hat q_i$ is fit to the data $X_1,\dots,X_{i-1}$  and is used to predict $X_i$, so that the sequence $\hat q_1, \hat q_2,\ldots$ progressively concentrates on whichever part of $\calH_1$ best explains the data. This is analogous to the role played by a Bayesian posterior, but arrived at by estimation rather than by conditioning on a prior}.

With Bayesianism in mind, let us give an example of a Bayes factor that is also an e-variable.

\begin{example}[t-test Bayes factor]\label{ex:ttest}
Consider iid observations $X_1,\dots,X_t \sim N(\mu,\sigma^2)$, where $\sigma>0$ is unknown, and suppose we wish to test the null $\calH_0 := \{N(0,\sigma^2) : \sigma>0\}$
against alternatives with nonzero standardized effect size $\delta := \mu/\sigma$. Let $W$ be a proper prior on $\delta$, and equip the nuisance scale $\sigma$ with the (improper) right-Haar/Jeffreys prior $w_H(\sigma)\propto 1/\sigma$. The resulting Bayes factor is~\citep{rouder2009bayesian}:
\begin{equation}
\label{eq:rh-bf}
B_t
:=
\frac{\int_{\mathbb R}\int_0^\infty \prod_{i=1}^t \frac{1}{\sigma}\rho\left(\frac{X_i-\sigma\delta}{\sigma}\right)\,\frac{\d\sigma}{\sigma}W(\d\delta)
}{\int_0^\infty \prod_{i=1}^t \frac{1}{\sigma}\rho\left(\frac{X_i}{\sigma}\right)\frac{\d\sigma}{\sigma}
},
\end{equation}
where $\rho$ denotes the standard normal density. Although the prior on $\sigma$ is improper, $B_t^{\mathrm{RH}}$ is an e-variable for $\calH_0$ (in fact, $(B_t)$ is an e-process for $\calH_0$)~\citep{grunwald2024safe,wang2025anytime}. Indeed, under a mild $2+\epsilon$-moment condition, this e-variable is log-optimal~\citep{perez2024statistics}.
\end{example}

To say more about the relationship of e-statistics to Bayes factors, if $\Theta_0=\{\theta_0\}$ is simple, then the Bayes factor is an e-variable; the logic is the same as in~\eqref{eq:lr-is-eval}. In general, however, the Bayes factor is not an e-variable: it only has expectation 1 under the prior-averaged null, not uniformly over every $\theta_0 \in \Theta_0$. That said,  \citet{grunwald2024safe} show that, for sufficiently regular parametric null hypotheses with simple alternatives, log-optimal e-variables are Bayes Factors under a specific prior on $\Theta_0$.

If the alternative hypothesis is composite, then one can obtain e-variables by equipping the alternative with an arbitrary prior determined by the analyst. There is then,
\revB{again under mild regularity conditions}, a unique prior on $\Theta_0$, depending on the prior on $\Theta_1$, such that the resulting Bayes factor is an e-variable.
In this way, optimal e-variables can be seen as a subset of Bayes factors, though only the prior on the alternative, not on the null, is independent of the analyst.

As for the relationship between likelihood ratios and Bayes factors, the latter reduces to the former in point versus point settings as the priors are determined (they are atoms). For composite hypotheses, the Bayes factor can be seen as an extension of the likelihood ratio if one allows for distributions over the parameter space.

Next let us turn to a powerful e-variable for bounded random variables.

\begin{example}[Bounded mean testing]
\label{ex:bounded-means}
For iid random variables $X_1,\dots,X_t$ lying in [0,1] with mean \revB{$\mu\in(0,1)$}, consider the object $M_t := \prod_{i=1}^t [1 + \lambda_i(X_i - \mu)]$, where $\lambda_i$ lies in $[-1/(1 - \mu), 1/\mu]$ and can be based on $X_1,\dots,X_{i-1}$ but not $X_j$ for $j\geq i$ (that is, it is \emph{predictable}). Then $M_t$ is nonnegative and satisfies $\E_\calH[M_t]=1$, where $\calH$ is the set of all distributions on [0,1] with mean $\mu$ (whether continuous, discrete, or mixed).
In fact, the process $(M_t)_{t\geq 0}$ is an e-process
for testing the null hypothesis that the data are iid from $P$ for any $P \in \calH$.
Such e-processes
were leveraged recently by \citet{waudby2024estimating} to obtain state-of-the-art confidence intervals/sequences.
\end{example}

The predictable sequence $(\lambda_t)_{t\geq 1}$ is typically chosen via the method of predictable plug-ins (see \citealt[Appendix B]{waudby2024estimating} for an overview), but the method of mixtures has also been studied in this context~\citep{stark2020sets}.

Note that unlike in universal inference, the example above defines an e-statistic for all distributions with mean $\mu$ without an explicit alternative in mind---similar to what we have seen in Example~\ref{ex:ryabko}. It is thus a good example of using a pragmatic, or instrumental, alternative to design a powerful e-variable for the null, which in this case is all distributions on $[0,1]$ with mean $\mu$. \rev{To elaborate, while no alternative is stated, one is nonetheless implicit. Indeed, every e-variable defines an implicit alternative. What varies from case to case then is not whether an alternative exists but whether it is one that should be considered as actually having generated the data, or simply something mathematically useful (e.g., Example~\ref{ex:ryabko}).}
Example~\ref{ex:bounded-means} also shows that in practice, e-variables may superficially look very different from likelihood ratios. Nevertheless, by  \citet[Proposition 4]{ramdas2020admissible}, it turns out that for all $P \in \calH$,
there exists some distribution $Q$ such that $M_t = q(X_1, \ldots, X_t)/p(X_1, \ldots, X_t)$, i.e., the likelihood ratio between $Q$ and $P$. Here the $q$ in the numerator varies along with the $p$ in the denominator. Such a ``likelihood-like'' interpretation holds for general e-statistics. We discuss this more in  Section~\ref{sec:counterarguments}.

It's worth emphasizing that all of the e-statistics discussed here obey the basic logic described in Section~\ref{sec:e-statistics} which lends them a unified evidential interpretation. This is unlike the situation with likelihood ratios, whose proposed generalizations do not sit comfortably together.

\section{Comparing Evidence Measures}
\label{sec:comparison}

We now spell out in more detail how p-values, Bayes factors, likelihood ratios, and e-statistics fare as measures of evidence. Table~\ref{tab:comparison} summarizes the discussion.

Regarding Bayes factors, we will draw a distinction between the objective and subjective traditions. In the objective tradition, priors are chosen by formal rules that are determined by the problem only and independent of the analyst~\citep{jeffreys1939theory,kass1996selection,berger2006case}, whereas in the subjective tradition priors can be any distribution. References to ``objective/subjective Bayes factors'' should be assumed to mean Bayes factors in the objective/subjective tradition.

We will also discuss both the \rev{simple-versus-simple} likelihood ratio, which we will refer to as simply the likelihood ratio, and the generalized likelihood ratio defined in~\eqref{eq:GLR}, which we will refer to as GLR.

For many of the desiderata listed in Section~\ref{sec:desiderata}, the distinction between e-variables and e-processes is immaterial. In these cases, we will refer simply to e-statistics without differentiating the two. The distinction becomes important when discussing counterfactuals and optional continuation; there we will be sure to explicitly distinguish them.

\subsection{Static Criteria}

\emph{Continuous measure.} P-values, likelihood ratios, and Bayes factors are all \revB{graded} measures of evidence \citep{schervish1996p,royall1997statistical}. Smaller p-values are interpreted as more evidence against the null; larger likelihood ratios and Bayes factors are interpreted as evidence in favor of $\Theta_1$ over $\Theta_0$. E-statistics are likewise \revB{graded}:  larger values are more evidence against the null. \revB{We stress that the criterion concerns the evidential scale rather than the topology of the map from data to evidence. For an event $A$ with $P(A)>0$, the indicator e-variable $E(X) = {\bf 1}_{A}/P(A)$ is itself  not a continuous function of $X$, yet the evidence it reports is `continuous' in the following sense: for any other e-value $E'$ relative to any other null hypothesis ${\cal H}'_0$, it holds that, if we observe  $E'(X) = E(X) + \epsilon$ for $\epsilon> 0$, then we are justified in saying that $E'$ provides more evidence against ${\cal H}_0$ than $E$ does against ${\cal H}_0$.}

\emph{Irrelevance of the analyst.}
The desideratum says that two analysts, analyzing the same data and the same $\calH_0$ (and, if present, $\calH_1$), should come to the same conclusions. This desideratum may be violated if (i) they employ different prior distributions within $\calH_0$ (and potentially $\calH_1$), whenever these are composite; (ii) they employ different test statistics; (iii) they employ different sampling plans (i.e., the rule that specifies how data will be collected) but happen to obtain the same statistic. Here we are concerned with whether the candidate notions satisfy the desideratum in the sense of not violating either (i) or (ii). (iii) is discussed further below in Section~\ref{sec:dynamic-criteria}.

It is easy to check that likelihood ratios and GLRs then satisfy the desideratum. Bayes factors in the subjective tradition do not satisfy it by design, whereas Bayes factors in the objective tradition do, as the prior is given by the structure of the problem. Intriguingly, while the ``subjectivity'' of Bayes factors is a common criticism of Bayesians made by frequentists, it's questionable whether p-values satisfy this criterion. Indeed, there are often multiple p-values for any given problem. Which one is calculated and reported is, of course, dependent on the analyst. We have thus marked p-values as only partially satisfying this requirement.

Similar comments apply to e-statistics.
There are multiple ways to play the betting game, and different analysts might thus use different e-statistics and come to different conclusions. Further, while they are perfectly well-defined in the frequentist setting, e-statistics \emph{can} be defined in the Bayesian setting and, as discussed above, optimal e-variables can often be recovered as Bayes factors with (sometimes peculiar) priors. Like p-values then, we mark e-statistics as only partially satisfying this  \revB{criterion}.

\emph{Coherence.}
 Both p-values and Bayes factors can be incoherent \citep{lavine1999bayes,fossaluza2017coherent}. Moreover, since the notion of \rev{coherence} requires evidence for or against \emph{composite} hypotheses, the ordinary \rev{simple-versus-simple} likelihood ratio is too narrow to address this desideratum.  The GLR, however, is coherent \citep{bickel2012strength,fossaluza2017coherent}.

Not all e-statistics are coherent.
\citet{bickel2024david} observed that the numeraire e-variable \citep{larsson2025numeraire} is not coherent, whereas the universal inference e-variable (and e-process) is. \rev{The upshot is thus not that e-statistics are coherent to some intermediate degree,  but that coherence is satisfied by some, but not all, e-statistics.} See \citet[Chapter 6]{ramdas2025hypothesis} for further discussion on coherent e-variables.

\emph{Consistency.}
Not all e-statistics are consistent, but many are.
In fact, it is typically easy to construct e-variables and e-processes which satisfy such a property. To give but a short list: those designed for testing means of bounded random variables \citep{waudby2024estimating}, those designed for testing exponential families~\citep{grunwald2024optimal,hao2024values}, two-sample testing \citep{shekhar2023nonparametric} and others \citep{waudby2025universal}. Still, there are obviously e-statistics that are inconsistent --- an example being the constant process $E_{\tau} \equiv 1$ that ignores all data. Meanwhile, recent work shows that universal inference can be conservative asymptotically, suggesting that universal inference e-values may not satisfy asymptotic consistency \citep{takatsu2025precise}.

The p-value is not consistent, while the likelihood ratio is \citep{grendar2012p}. The GLR is consistent under regularity and separability assumptions~\citep{bickel2012strength}, but not in general. Bayes factors are
consistent in most practical settings (so we marked the desideratum as satisfied), though there are  some exceptions
depending on the prior and the hypothesis class \citep{casella2009consistency,moreno2010consistency}.

\emph{Sample size invariance.}
Both likelihood ratios and subjective Bayes factors have sample size invariant interpretations. The likelihood ratio, for instance, is interpreted as:
\emph{if $p_{\theta_1}(X) / p_{\theta_0}(X)=c$, then the data are $c$ times as likely under $p_{\theta_1}$ as they are under $p_{\theta_0}$,} a statement which does not mention the sample size. Or, in the words of \citet{goodman1988evidence}, ``the
likelihood ratio has the same meaning in trials of different
designs and sizes.'' Similarly for the GLR. Meanwhile, e-statistics
can be seen as monetary gains in a betting game, an interpretation
which also doesn’t depend on the sample size. P-values, on the other hand, are not sample size invariant~\citep{goodman1988evidence}.
For objective Bayes, the answer depends on the hypotheses under consideration. For some hypotheses (such as in linear regression), the prior advocated in some objective
Bayes methods depends on the covariates \citep{de2021optional} and therefore, implicitly, also on the sample size. Thus, sample size invariance fails.

\begin{table*}[p]
\renewcommand{\arraystretch}{1.19}
\centering
\small
\begin{tabular}{@{}l r | p{1.1cm} | p{0.6cm} p{0.6cm} | p{0.6cm} p{0.6cm} | p{0.6cm} p{0.6cm}@{}}
\emph{Desideratum} & \emph{Trad.}
& \emph{p-values}
& \multicolumn{2}{c|}{\emph{LRs}}
& \multicolumn{2}{c|}{\emph{Bayes factors}}
& \multicolumn{2}{c}{\emph{e-statistics}} \\
&&& simp. & GLR & subj. & obj. & all & some \\
\hline
Scalar measure
& B,E,L
& $\checkmark$
& $\checkmark$
& $\checkmark$
& $\checkmark$
& $\checkmark$
& $\checkmark$
& $\checkmark$
\\
Continuous measure
& B,E,L
& \checkmark
& \checkmark
& \checkmark
& \checkmark
& \checkmark
& \checkmark
& \checkmark \\
Irrelevance of analyst
& E,L
& $\approx$
& \checkmark
& \checkmark
& -
& \checkmark
& $\approx$
& \checkmark \\
Coherence
& O
& -
& -
& \checkmark
& -
& -
& -
& \checkmark \\
Consistency
& B,L
& -
& \checkmark
& $\approx$
& \checkmark
& \checkmark
& -
& \checkmark \\
Sample size invariance
& B,L
& -
& \checkmark
& \checkmark
& \checkmark
& $\approx$
& \checkmark
& \checkmark \\
Scale invariance
& B,L
& -
& \checkmark
& \checkmark
& \checkmark
& \checkmark
& $\approx$
& \checkmark \\
Reparam.\ invariance
& B,L
& -
& \checkmark
& \checkmark
& $\approx$
& $\approx$
& $\approx$
& \checkmark \\
Single hypothesis validity
& E
& \checkmark
& -
& -
& -
& -
& \checkmark
& \checkmark \\
Error rates
& E
& \checkmark
& \checkmark
& $\approx$
& -
& -
& \checkmark
& \checkmark  \\
Composite hypotheses
& O
& \checkmark
& -
& \checkmark
& \checkmark
& \checkmark
& \checkmark
& \checkmark \\
Nonparametric hypotheses
& O
& \checkmark
& -
& \checkmark
& $\approx$
& $\approx$
& \checkmark
& \checkmark \\
Likelihood principle (Strict)
& L
& -
& \checkmark
& \checkmark
& \checkmark
& -
& -
& $\approx$  \\
Likelihood principle (Loose)
& L
& -
& \checkmark
& \checkmark
& \checkmark
& $\approx$
& \checkmark
& \checkmark \\
Composability
& O
& \checkmark
& -
& -
& -
& -
& \checkmark
& \checkmark\\
Accumulation I (Fixed)
& B,E,L
& \checkmark$^{+}$
& \checkmark$^{+}$
& -
& \checkmark
& $\approx$
& \checkmark$^{+}$
& \checkmark$^{+}$
\\
Accumulation II (Flexible)
& O
& -
& \checkmark$^{+}$
& -
& \checkmark
& -
& \checkmark$^{+}$
& \checkmark$^{+}$
\\
Dynamic Consistency
& B, O
& -
& \checkmark$^{+}$
& -
& \checkmark
& -
& -
& \checkmark$^{+}$
\\
Inter-experiment counterfactuals
& O
& -
& \checkmark$^{+}$
& -
& \checkmark
& -
& \checkmark$^{+}$
& \checkmark$^{+}$
\\
Intra-experiment counterfactuals
& B,L
& -
& \checkmark$^{+}$
& -$^*$
& \checkmark
& -
& -
& $\approx$ \\
Intra-experiment optional stopping
& B,L
& -
& \checkmark$^{+}$
& -$^*$
& \checkmark
& -
& -
& \checkmark$^{+}$
\\
\end{tabular}
\caption{Comparison of how different evidence functions fare with respect to various evidential desiderata.
Here $\checkmark$ indicates the desideratum is largely satisfied, ``--'' that it is either not satisfied or cannot be meaningfully defined, ``$\approx$'' that it is partially satisfied.
In the final six lines, $\checkmark^+$ means that the evidence is well-defined, has a clear interpretation and provides valid error rates. $\checkmark$ means that it  is well-defined and has a clear interpretation, but in general cannot be used to infer valid error rates. `--$^*$' means that the evidence remains  well-defined\revB{---and, in the case of intra-experiment counterfactuals, numerically unchanged---}so the desideratum is formally satisfied, yet it lacks any clear justification and does not lead to valid error rates, so it is hard to imagine that one would ever want to use it.
The ``Trad.''\ column indicates which tradition the desideratum comes from:  L = likelihood tradition, B = Bayesian tradition, E = error-probability tradition, O = Other.
Since there are typically multiple e-statistics for a given hypothesis, we break their analysis into two categories: `all' meaning all e-statistics which satisfy the criteria, or `some' meaning that there exists at least one e-statistic which does (though typically there are many).
See the text for more discussion in each case. }
\label{tab:comparison}
\end{table*}

\emph{Scale invariance.}
Both likelihood ratios and Bayes factors are invariant to one-to-one transformations of the data because they are ratios of probabilities. To elaborate, under any such transformation both densities are multiplied by the same quantity: If $Y = g(X)$ then the new density for $Y$ is $p_{\theta_j}(y) = p_{\theta_j}(x) |\det J_{g^{-1}}(y)|$ where $J_{g^{-1}}$ is the Jacobian.
Thus, the change to both numerator and denominator cancel out. (Note the Jacobian only appears in continuous-valued data; for discrete data the invariance is immediate.) The GLR is scale invariant for the same reason.

Scale invariance for e-statistics is more subtle. Some e-statistics automatically satisfy this property, such as universal inference (again using that it is the ratio of probabilities) and those based on self-normalized processes \citep{wang2025anytime} (since self-normalized statistics are scale free).

However, not all e-statistics are immediately scale invariant. Consider $E(X) = \mathbf{1}\{X\leq 1\} / P(X\leq 1)$ for any distribution $P$ with $P(X\leq 1)>0$. This is clearly an e-variable. If $Y = aX$, then $E(Y) \neq E(X)$, so $E$ is not scale invariant. This is because a threshold has been hard-coded in the example. One should arguably parameterize this threshold and instead consider the family $E_s(X) = \mathbf{1}\{ X\leq s\}/P(X\leq s)$. In this case, $E_s(Y) = E_{s/a}(X)$.

Allowing the parameter space to change alongside the data is natural. For instance, following Example~\ref{ex:subgaussian}, $E_\lambda(X) = \exp(\lambda X - \lambda^2/2)$ is an e-variable for $\calH=\{ N(0,1)\}$. Suppose we let $ Y = aX$, so under the null, $Y\sim N(0,a^2)$. Then $E$ is scale invariant in the sense that \revB{$E_\lambda(X) = E_{\lambda/a}(Y)$, where on the right $E$ denotes the family of Example~\ref{ex:subgaussian} with null variance $a^2$}. Notice that $E_\lambda(X)$ is the likelihood ratio of $N(\lambda,1)$ and $N(0,1)$. That is, the parameter $\lambda$ is encoding the alternative distribution. Hence, when we refer to the likelihood ratio being scale invariant, we are implicitly allowing the parameter to change.

\emph{Reparameterization invariance.}
Using the same logic as above, likelihood ratios are invariant to transformations of parameters (cf.\ \citealt{mccullagh1986invariants}). So too are Bayes factors if one allows the prior to change along with the reparameterization. That is, if we begin with a prior $\pi$ over parameters $\theta$, reparameterize $\theta \mapsto \phi(\theta)$, and consider the pushforward measure $f(\phi) := \pi(\theta)|\det \d\theta/\d\phi|$ as the new prior, then the Bayes factor is invariant.

However, considering the pushforward is not always done in practice. For instance, choosing a uniform prior in one parameterization will not necessarily yield a uniform prior in another. Thus, if one is committed to a specific class of priors, the Bayes factor can depend on the particular parameterization. Some Bayesians find this troubling, which led to  invariant priors \citep{jeffreys1946invariant}. Neither objective Bayes factors nor subjective Bayes factors always use invariant priors, however.

A similar nuance applies to e-statistics as it did in the case of scale invariance. As an object satisfying $\sup_{P\in\calH}\E_P[E]\leq 1$, a change in the parameter space simply relabels the null hypothesis, and does not affect the guarantee afforded by the e-statistic. In other words, e-statistics are invariant to transformations of the parameter in the sense that they remain e-statistics. However, if one is committed to a specific ``generating procedure'' for e-variables (e.g., universal inference with prior $\pi(\theta)$, or constructing $E_\lambda$ using a specific $\lambda$), then its value might change after a transformation of the parameters. Therefore, we mark general e-statistics as only partially satisfying this requirement.

Finally, not all p-values are reparameterization invariant, leading to the search for some which are~\citep{evans2010invariant}.

\emph{Single hypothesis validity.}
P-values and e-statistics are defined only in terms of a single hypothesis $\calH$, whereas both likelihood ratios and Bayes factors require two. See, e.g., Example~\ref{ex:ryabko} and~\ref{ex:bounded-means}. That is, the latter provide relative evidence between two hypotheses and the former provide a notion of absolute evidence against a single hypothesis.

\emph{Long-run error rates.}
This is perhaps the defining feature of evidence from the error-probability tradition. It holds that if one uses a measure of evidence to make decisions (which is itself already a controversial endeavor, and marks one of the divides between the Fisherian and Neyman-Pearson schools of thought), then the frequentist error on those decisions should be controlled. More precisely, if we repeat the decision process many times, we should have some guarantee on how often the decision will be wrong.

As discussed in Section~\ref{sec:e-statistics}, Markov's inequality (and Ville's inequality for e-processes) gives us control over type-I error rates for e-statistics. The likelihood ratio, being an e-value, also gives error control, but the GLR on its own does not. There, error rates must be obtained by appealing to the sampling plan, underlying distribution, or other machinery.
Since in practice, GLRs are almost always used within a context in which such error rates are analyzed as function of the GLR for a given sampling distribution, we still marked them as approximately satisfying this criterion.
Subjective Bayes factors also do not provide type-I error control~\citep{grunwald2024safe}; for objective Bayes factors, there are some cases in which type-I error control is provided, such as the t-test Bayes factor of Example~\ref{ex:ttest}, but usually it is not (note that we refer to exact, nonasymptotic error rates here).

It's worth mentioning so-called ``post-hoc validity,'' a new development in the theory of e-statistics.
When measures of evidence are used to make decisions,
e-statistics continue to provide meaningful error guarantees, in the form of control on expected risk, under data-dependent significance levels \citep{grunwald2023posterior,grunwald2024beyond,chugg2026admissibility}. This is not allowed by traditional p-values. \rev{The property matters because significance levels are, in practice, often not fixed before the data are seen. An analyst who sets out with $\alpha = 0.05$ but, on seeing a striking result, reports it at $\alpha = 0.01$ has broken the contract under which the p-value's risk guarantee was issued.  Post-hoc validity says that the guarantee attached to an e-value survives such after-the-fact choices of level.} Further, the only subset of p-values which do provide such a guarantee turn out to be precisely the inverse of e-variables, both in asymptotic and non-asymptotic settings \citep{wang2022false,koning2023post,chugg2026post}.

\emph{Composite and nonparametric hypotheses.}
The only evidential measure that, by definition, cannot handle composite hypotheses is the \rev{simple-versus-simple} likelihood ratio. All other measures handle composite hypotheses and, in principle, even handle highly complex nonparametric hypotheses.

Some evidential measures, however, handle such complex classes more naturally than others. For instance, there exist simple e-statistics for the class of all bounded distributions with mean $\mu$ (Example~\ref{ex:bounded-means}), which is a large nonparametric family. By contrast, to define a Bayes factor for this same hypothesis one must place a prior on an infinite-dimensional space of distributions (or otherwise restrict attention to a parametric or semiparametric sub-model). This can certainly be done, but it is considerably less direct and substantially more prior-dependent (and can lead to less evidence against hypotheses that are evidently wrong in light of the data; see \citealp{li2023lilidaiscontribution}). For this reason, we regard this criterion as only partially satisfied by both subjective and objective Bayes factors.

\emph{The likelihood principle.}
The likelihood principle is something like the founding charter, or the constitution, of the likelihoodists.
It is useful to distinguish between two versions: the  \emph{strict} and \emph{loose} likelihood principle.
The strict version holds that if two data sets induce proportional likelihood functions for the parameters of interest (i.e., for all parameters in $\calH_0$ and $\calH_1$), then the evidence should be the same for both data sets. The loose version holds that evidence can always be written as, or at least upper bounded by, a likelihood ratio with some element of $\calH_0$ in the denominator.

Let us deal first with the strict likelihood principle. This is, of course, satisfied by the likelihood ratio.
For the subjective Bayes factor, it is satisfied if one identifies likelihood with marginal likelihood over the prior.
Here we assume, as is usually done, that the priors are chosen \emph{independently} of the sampling plan, i.e. the analyst's prior assessment of the parameters determining the actual distribution is independent of the sampling plan employed. This is in contrast to objective Bayesian approaches, in which priors are chosen by formal means that often depend on aspects of the sampling plan such as the stopping time \citep{berger1988likelihood}. Indeed,  \citet{berger1988likelihood,de2021optional} give  concrete examples where Jeffreys' prior depends on the sampling plan, hence the value of the objective Bayes factor is not merely determined by the likelihoods and as such it violates the strict likelihood principle. P-values violate it for the same reason, whereas the GLR satisfies it.

As for e-processes, the strict likelihood principle is satisfied by the likelihood ratio e-process and the universal inference e-process, \revB{when the alternative is specified by a prior as in Example~\ref{ex:universal-inference}, Eq. \eqref{eq:chatcatch}}. 
%(Example~\ref{ex:universal-inference}).
There is a subtle catch which prevents us from saying that the strict version is always satisfied by e-statistics
\revB{of the form \eqref{eq:chatcatch}}, however. We describe this issue further below under intra-experiment counterfactuals.

The loose version of the likelihood principle is clearly satisfied by any evidential measure satisfying the strict version. Moreover, it is once again violated by p-values.  As to e-statistics, as shown in \citet[Theorem 14.5]{ramdas2025hypothesis} (a special case of which we mentioned in Section~\ref{sec:examples}), any e-variable $E$ for $\calH$ can be shown to be dominated by likelihood ratios, in the sense that for any $P\in \calH$, there exists some $Q$ such that $ E(X) \leq q/p$; see Section~\ref{sec:counterarguments} for a simple proof. For objective Bayes, this is not the case, but since objective Bayesian evidence can always be written as a likelihood ratio with a (potentially improper) prior over $\Theta_0$ in the denominator, we still mark it as partially satisfied.

\emph{Composability under dependence.}
The convex combination of any set of (arbitrarily dependent) e-statistics remains an e-statistic: If $E^1, \dots,E^K$ are e-variables then 
\revB{$\lambda_0 + \sum_{1 \leq k\leq K}\lambda_k E^k$
is an e-variable where $\sum_{0 \leq i\leq K}\lambda_i \leq 1$,}
$\lambda_i \geq 0$. Similarly for e-processes.  In fact, the only admissible way of merging e-variables is with a combination such that $\sum_{0 \leq i\leq K}\lambda_i=1$~\citep{wang2025only}.
There also exist rules for combining arbitrarily dependent p-values, including two times the average and the median of the p-values. See \citet{vovk2022admissible} for a nice overview.
\revB{Marginal likelihood-ratios or Bayes-factors from separate sources cannot in general be merged when their dependence structure is unknown.}
%Bayes factors and likelihood ratios are not composable under dependence; they require (conditional) independence---see the first desideratum in the following section.

\subsection{Dynamic Criteria}
\label{sec:dynamic-criteria}

To assess the dynamic criteria introduced in Section~\ref{sec:desiderata}, it will be convenient to introduce some unified notation that we can use throughout the section. We will assume that we want to measure the evidence of some data $X^*$ which decomposes as $X^* = (X_{(1)}, X_{(2)},\dots,X_{(\tau)})$.
For the inter-experiment dynamic criteria (the first four criteria below), $X_{(i)}$ is itself the data produced by some study. For the intra-experiment criteria, $X_{(i)}$ is more fruitfully considered to be a single datum and $X^*$ are the data of one study. In both scenarios, the total number of studies run or observations collected, $\tau$, may or may not be known in advance (depending on the desideratum). \\

\emph{Accumulation I: Fixed-Design.}
If the $X_{(j)}$ are independent and $\tau$ is known in advance or determined independently of the data, then the likelihood ratio decomposes as $\lr = \prod_{j\leq \tau} p_1(X_{(j)})/p_0(X_{(j)})$, meaning that we can combine evidence by multiplication.

As is well known, the same multiplicative accumulation holds for Bayes factors when the priors $\pi_0$ and $\pi_1$ are fixed in advance, independently of the data-collection process (what we call the subjective Bayesian setting). In that case, independent studies can be combined sequentially by updating each prior to the corresponding posterior after each study. The overall Bayes factor is then the product of the stage-wise Bayes factors.

The situation is less clear for objective Bayes methods. There, the prior is often not a genuinely fixed ingredient, but is instead chosen by a rule that may depend on features of the experimental design, such as the sample space, stopping plan, or covariate structure. If later studies are added, the prior deemed appropriate for the combined experiment may differ from the prior used for the first study considered in isolation. As a result, sequential accumulation becomes ambiguous: the product of the stagewise Bayes factors need not coincide with the Bayes factor that would have been computed had the full experiment been specified from the start \citep{de2021optional}. Since this happens only for some, and not nearly all, choices of $\calH$, we marked the requirement as only partially satisfied.

The GLR, meanwhile, is not additive on any scale, as the parameter achieving (or approximating) the supremum may change for different samples.
\revB{Independent p-values can be combined using Fisher's method, which adds not the p-values themselves but their negative logarithms}, and
the product of two conditionally independent e-statistics remains an e-statistic, a property which follows easily from the definition of the expected value. See \citet[Chapter 8]{ramdas2025hypothesis} for more discussion.

\emph{Accumulation II: Flexible-Design.}
Now suppose that the decision to perform an additional study depends on the results of previous studies.
If the data $X_{(j)}$ is independent of $X_{(1)},\dots,X_{(j-1)}$ and $X_{(i)}$ is associated with e-variable $E_i$, then the product $\prod_{i\leq \tau} E_i$ is an e-variable.
This was a major motivation in the original development of e-statistics~\citep{grunwald2024safe}. This property is also satisfied by e-processes that, in the inter-experimental setting, are constructed on the fly from a sequence of e-variables.

GLRs fail to satisfy Accumulation II for the same reason they failed to satisfy Accumulation I. Objective Bayes factors fail this criterion more dramatically than they did Accumulation I: the prior for the initial experiment may not only be chosen in different ways, its definition will in general \rev{depend} on the number of studies that will eventually be done. Hence the objective Bayes factor may in fact be undefined.
Subjective Bayes factors, on the other hand, continue to satisfy this criterion~\citep{rouder2014optional,de2021optional}.

P-values fail for this and all subsequent desiderata because they require that $X^*$ is fully specified in advance of running the study. Consider the following example.

\begin{example}\label{ex:oc}
Suppose that a randomized clinical trial to test a new medical treatment is performed on 50 patients represented by 50-dimensional data vector $X_{(1)}$. The result turns out to be {\em promising but not conclusive}: the researchers  observed a p-value of $p_1 = 0.1$ while they had a significance level of $0.05$ in mind. But their boss is optimistic at the news and agrees to supply the resources to test another 30 patients, resulting in data $X_{(2)}$.

Is this good news? Not if one insists on measuring evidence in the second trial by another p-value, say $p_2$. 
% (see below)
\revB{For some realizations of $X_{(1)}$, we might decide not to gather $X_{(2)}$. As a result, standard combination methods for p-values like Fisher's cannot be employed any more, since they invariably require that, no matter what is observed in each sample, we always combine both.
}
%The decision to gather  $X_{(2)}$ depends on $X_{(1)}$ and, as a result, combination methods like Fisher's cannot be employed any more. Indeed, they require 
%that $X^*= (X_{(1)}, X_{(2)})$ whereas in our setting, $X^*$ would be equal to $X_{(1)}$ for some values of $X_{(1)}$ and equal to  $(X_{(1)},X_{(2)})$ otherwise.
Similarly, joining the two data sets and recalculating the p-value leads to a wrong answer as well. 
\revB{
Indeed, take a method that stops for some values of $X_{(1)}$ and in that case outputs $p^*= p_1$, a sharp, nonconservative p-value for $X_{(1)}$, yet continues for other values of $X_{(1)}$ and in that case, after observing $X_{(2)}$, outputs $p^*=p'$ with $p'$ a number strictly smaller than $1$. It is easily shown that for any such method, the resulting $p^*$ is not a valid p-value anymore.
}
In fact, the only known way to obtain a sequence of \emph{anytime-valid} p-values in such settings is to convert each $p_i$, $i=1,2$, into an e-value $E_i$ by calibration (Section~\ref{sec:examples}), and then report not $p_i$ itself but $p_i' := 1/E_i$. The quantity $p'_{12} := 1/(E_1E_2)$ can then be interpreted as a p-value for the combined data. In practice, however, this approach is typically inefficient \citep[Section 7]{grunwald2024safe}, and one can do better by working directly with e-values.
\end{example}

\emph{Dynamic Consistency.}
Dynamic consistency asks that we obtain the same evidence when we treat $X^*$ as a batch versus  when we
treat samples $X_{(j)}$ separately, sequentially, and then combine them.
For p-values, likelihood ratios, and Bayes factors,
satisfying dynamic consistency is equivalent to satisfying accumulation under a flexible design (here we again assume that $\tau$ can be data-dependent).

For e-statistics we must distinguish between several cases. If we design an e-process for the stream of data $X_{(1)},X_{(2)},\dots$, then dynamic consistency is satisfied~\citep{ramdas2023game}.
If, however, we model the individual $X_{(j)}$ by separate e-variables and combine them via multiplication, then this desideratum is not satisfied in general \citep[Section 6]{grunwald2024safe}. In particular, for some e-variables the product $\prod_{j=1}^{\tau} E_j$ of the e-values $E_j$ for study $X_{(j)}$ does not coincide with the single e-value $E^*$ for $X^*$ that one would have obtained if one had designed a log-optimal e-variable for $X^*$ directly (though we emphasize that, as stated earlier, the product $\prod_{j=1}^\tau E_j$ is a valid e-variable, which of course yields valid error rates as discussed in Section~\ref{sec:e-statistics}).

\emph{Inter-experiment counterfactuals.}
A common criticism of p-values is that, paradoxically, they depend on data that are not observed, or, more precisely, on what decisions an experimenter would have taken in counterfactual situations in which the data were different~\citep{barnard1947meaning,wagenmakers2007practical}. There are multiple kinds of counterfactuals to consider, which we delineate as inter- and intra-experiment counterfactuals\footnote{We have not found this distinction anywhere else in the literature, but it is important to make since the criteria behave quite differently under each sub-division!}. The former, discussed here, concerns counterfactuals with respect to the number of observations, which might also be thought of as optional continuation in counterfactual situations.

\begin{example}[Example~\ref{ex:oc}, continued]
\label{ex:ocb}
In a variation of the previous example, suppose the researchers told their boss merely that the p-value was small enough for the result to be ``promising but not conclusive,'' and not its actual value. The boss, once again, responds optimistically and requests that they continue the trial on a second batch of 30 patients.
The researchers, who know their statistics, are now worried about invalidating the results. But then news reaches the boss that the p-value is 0.1. Dismayed, he decides to stop the trial.

Should the researchers be relieved? {\em The answer is an emphatic no\/}: a calculation \revB{similar to the one in the previous example} shows that, \revB{if the originally reported p-value was sharp}, then the mere fact that the sample would have been $80$ patients (thus different from the originally planned $50$) in a counterfactual situation
(i.e. if the first 50 data points had been different than they actually were) makes the p-value invalid. That is, counterfactuals can ruin the validity of the p-value even if the sample plan was not in fact changed for the data which were actually observed.
\end{example}

The dependence of the p-value on decisions that would have been made in counterfactual situations is disturbing, since in reality, it is often quite unknowable (or even meaningless to speak about) what exactly would have happened had the data been different from what they actually were. That p-values cannot handle counterfactuals is an immediate consequence of the fact that they cannot deal with accumulation. Similarly for the objective Bayes factor and the GLR.

For e-variables and e-processes in inter-experimental settings (where the e-processes are constructed by multiplying the e-variables), this desideratum is satisfied because the constructed e-variables are valid at arbitrary stopping times. That is, regardless of the data witnessed thus far, stopping an e-process will always result in a valid e-value. Similarly for subjective Bayes factors, which also remain valid at stopping times.

Note that this implies that p-values which can be written as the inverse of e-values also satisfy this property. In general, however, such p-values are conservative and are not the ones used in practice.

\emph{Intra-experiment optional stopping.}
Now we suppose that $X^*$ represents data from a single study in which we observe $\tau$ data points, $X^* = (X_1, \ldots, X_{\tau})$. As usual, $\tau$ is not fixed: it is a stopping time whose definition the analyst may not know or may not want to fix in advance. Consider the following simple example in the spirit of \citet{lindley1976inference}.

\begin{example}
\label{ex:optional_stop}
Suppose the data are binary and we observe $X^*=(0,1,0,1,1)$, but do not know whether the sample size was fixed to be 5 in advance or if the sampling plan was ``stop as soon as you see two $1$s in a row.'' Or, then again, maybe each data point is expensive and  the $p$-value for a sample size of $5$ was already so small that your boss decided data collection should be stopped.
\end{example}

As we stated in Section~\ref{sec:desiderata}, this criterion is very similar to Accumulation II, but  making the distinction between optional continuation
`inside' a single study versus   `between' studies sheds light on several subtleties for e-statistics.
As we saw in Section~\ref{sec:e-statistics}, e-processes are defined such that they allow optional stopping. Likelihood ratios, being e-processes, also satisfy this criterion. On the other hand, for e-variables (when directly defined on the batch $X^*$, as they often are in practice), optional stopping is an undefined operation.

As for the other notions of evidence, it is well-known that p-values do not allow for data-dependent stopping times. In fact, picking the sample size as a function of the data and reporting the resulting p-value has come to be known as ``p-hacking,'' and is often cited among the most common ``questionable research practices'' \citep{simmons2011false,john2012measuring}.
Bayes factors, meanwhile, allow for optional stopping as long as the priors are chosen a priori, independently of the stopping time (though note that they do not in general provide type-I error rates, in contrast to likelihood ratios).
We refer to \citet{de2021optional} and \citet{rouder2014optional} for more discussion on Bayes factors and optional stopping.

Like the Bayes factor with subjective prior, the GLR remains well-defined under optional stopping, but like the Bayes factor, it fails to provide valid error rates, as the supremum can exaggerate evidence for the alternative~\citep{ramdas2023game}. Therefore, while the Bayes factor still has a clear interpretation and can be justified in case the priors reflect trustworthy prior knowledge, using the GLR here has no clear justification at all. In practice, the GLR is usually  employed within the error-probability tradition, and its main goal is to provide error rates, which it cannot provide here.

\emph{Intra-experiment counterfactuals.}
Researchers within the likelihood tradition have observed that the p-values' dependence on decisions made in counterfactual situations stretches beyond only the size of the dataset, as illustrated by the following celebrated example:

\begin{example}[\textbf{Pratt's voltmeter}, cf.\
\citealp{edwards1972likelihood,savage1962foundations}]
Suppose we observe $X_1, \ldots, X_n$ where $n$ is fixed and the $X_i$ represent voltages of electron tubes,  measured with an accurate voltmeter.
    A statistician  examines the  $X_i$ assuming they are normally distributed with fixed variance and some mean $\mu$. He aims to use a $p$-value to measure the evidence against the null hypothesis that $\mu=7.0$.
Later he visits the engineer’s laboratory, and notices that the voltmeter reads only as far as 10: the population appears to be {\em censored}. Even though none of the $X_i$ were $10$, this makes the standard $p$-value invalid; it necessitates a new calculation based on a $p$-value that takes into account the (potential, counterfactual) censoring.
However, the engineer says she also has a super-high-range-meter, equally accurate, which she would have used if any of the measurements had  turned out $\geq 10$.
This is a relief to the statistician, because it means the  original $p$-value is correct after all.
But the next day the engineer telephones and says, {\em “I just discovered my high-range voltmeter was not working the day I did the experiment”}.
The statistician then informs her that a new analysis will be required after all!
The engineer is astounded. She says, ``But the experiment turned out just the same as if the high-range meter had been working. {\em I learned exactly what I would have learned if the high-range meter had been available}. Next you’ll be asking about my oscilloscope!''
\end{example}

We may think of both the voltmeter example and Example~\ref{ex:optional_stop} as an instance of the following generic phenomenon.

The random variable $X^*$, taking values in some set ${\cal X}^*$, defines a partition of the sample space $\Omega$, say $\{ \cE_{x} : x\in  {\cal X}^*\}$, where $\cE_{x}$ is the set of exactly those $\omega \in \Omega$ with \revB{$X^*(\omega)= x$}.  Observing $X^*=x$ is equivalent to observing that the event $\cE_x$ occurred. Now, it is usually tacitly assumed that the definition of $X^*$ is known to the observer. In practice, however, an analyst simply observes an event $\cE_x$, for some $x\in {\cal X}^*$. The analyst then knows that $\cE_x$ happened but often does not know how $X^*$ was even defined. This is the case in Example~\ref{ex:optional_stop}, where we observed
$X^*=(X_1, \ldots, X_{\tau}) = (0,1,0,1,1)$ but did not know whether $\tau$ was defined as the constant $5$ or  as `stop when you see two $1$s'. It is also the case in the voltmeter example, where we observed $(x_1,\ldots, x_n)$ with all $x_i < 10$ but do not know whether
$X^*= (X_1, \ldots, X_n)$ or $X^*= (X_1 \wedge 10, \ldots, X_n \wedge 10)$.

Now, above we defined likelihood ratios as functions of the random variable $X^*$, but we may equivalently define them as functions of events, i.e. measurable subsets of $\Omega$. To avoid difficult mathematics, let us illustrate this in the case of discrete data.  For any event $\cE$, any random variable $X$, and any value $x$ such that $X=x$ corresponds to event $\cE_x$, we have $p_1(x)/p_0(x) = P_1(\cE)/P_0(\cE)$ as long as $\cE= \cE_x$. Thus, we can re-define likelihood ratios as functions of events and will get the same results as before, irrespective of the definition of $X^*$. As a consequence, the likelihood ratio is immune to decisions in counterfactual situations in a very general sense. The same holds for the subjective Bayes factor. \revB{For the objective Bayes factor and isolated e-variables, the desideratum fails to be met for the same reasons as for inter-experiment counterfactuals. The GLR is a different case. Being a functional of the likelihood alone, its value is unchanged under such counterfactuals. In the voltmeter example, if every observation falls below the censoring point, the likelihood---and hence the GLR---is exactly what it would have been had the voltmeter been unrestricted. What does change is its sampling distribution, so thresholds calibrated under one protocol need not retain their error guarantees under the other. Table~\ref{tab:comparison} thus records `--$^*$' for the GLR. }

The situation is trickier for e-processes. Since likelihood ratios define e-processes, we might say this desideratum holds for at least some e-processes. On the other hand, one invariably defines e-processes on filtrations (essentially sequences of random variables), not on arbitrary events, and at the time of writing it is not clear whether they can be extended to events beyond the simplest case in which ${\cal H}_0$ and ${\cal H}_1$ are simple (this is also the reason we only marked the strict likelihood principle to be only approximately satisfied by e-processes --- likelihood ratios are defined on arbitrary events, and e-processes on the more restricted notion of filtrations). Further work is required to answer this question.  \\

Overall, we find that while no single e-statistic satisfies all evidential desiderata in Section~\ref{sec:desiderata}, each desideratum is (at least partially) satisfied by at least one e-statistic. Moreover, when we restrict attention to criteria satisfied by \emph{all} e-variables, they satisfy the largest set of desiderata among those measures which can handle composite hypotheses (which we deem to be a very important property!).
Indeed, as we've \revB{emphasized} throughout and will explore more shortly, we may think of e-statistics as providing a novel extension to likelihood ratios, arguably more sophisticated than the generalized likelihood ratio, that behaves particularly well on most common desiderata.

With that said, let us not pretend that e-statistics are the perfect measure of evidence. We end by considering some of their drawbacks.

\section{Limitations and Counterarguments}
\label{sec:counterarguments}

One could of course object to e-statistics as a measure of evidence by appealing to any of the criteria listed above that they do not wholly satisfy. Instead, in this section we consider two broader critiques.

\paragraph{Non-uniqueness.} The first is the fact that there can be many e-variables for a given problem. Even if one uses e-statistics, the amount of evidence against a hypothesis is not determined by the structure of the problem alone, but \rev{also by} how the analyst chooses to bet in the betting game.

We are sympathetic to this concern, but let us point out that the situation is not so different from other measures of evidence. For instance, there are typically multiple p-values that one may use for a given problem, and the Bayes factor depends on the choice of prior which may vary from analyst to analyst. Further, while the likelihood ratio for simple hypotheses is uniquely determined,  when moving to composite settings the analyst has \rev{their} choice of possible generalizations.\footnote{\rev{One might also tie this non-uniqueness to a familiar epistemological point, which is that data are always interpreted through our theories, and there is thus never only one way to interpret them.  In the jargon, observation is always \emph{theory-laden}~\citep{popper1963conjectures}.}}

Nevertheless, the situation is arguably somewhat clearer for p-values than for e-values. In particular, within the error-statistics tradition, there is broad consensus that the quality of a p-value, when used for testing, is determined by the maximal power it can achieve. For sufficiently simple $\calH_0$ and $\calH_1$---though only in such cases---one can define a uniformly most powerful p-value. In the e-value literature, by contrast, the preferred optimality criterion under a simple alternative is often taken to be log-optimality. For composite alternatives, however, there are multiple ways of generalizing log-optimality, and these lead to quite different e-statistics. More broadly, there is less consensus in the e-statistics literature that log-optimality is the right criterion than there is in the error-statistics tradition that power is the right one \citep{koning2024continuous}.

\paragraph{Birnbaum's theorem and the likelihood principle.}
\citet{birnbaum1962foundations} showed that two conditions entail the likelihood principle: the sufficiency principle and the conditionality principle. The result has come to be known as Birnbaum's theorem.

The sufficiency principle states that if there exists a sufficient statistic for the statistical model, then it captures everything about evidence. That is, if two experiments give the same sufficient statistic, then they provide the same evidence for the parameter.
The conditionality principle states that evidence should only depend on which experiment was actually run. For instance, if an analyst decides between two experiments by flipping a coin, the evidence should depend only on that which was performed.

Birnbaum's theorem is controversial because the conditionality and sufficiency principles are relatively weak, striking many frequentists as perfectly plausible. The conclusion, however,
contradicts the frequentist's favorite inferential methods. The theorem has thus fallen under various forms of scrutiny \citep{durbin1970birnbaum,kalbfleisch1975sufficiency,evans1986principles,evans2013does,gandenberger2015new,mayo2014birnbaum}.

To make the controversy more concrete, consider again Example~\ref{ex:optional_stop}.
The likelihood ratio is entirely determined from these data, thus the likelihoodist  is satisfied that they have enough evidence about the situation to make a pronouncement. But the frequentist who wants to employ a p-value is not. For them, the sampling plan matters: the p-value differs in all cases.

How do e-statistics fit into this story? As we've emphasized throughout the article, e-statistics are intimately related to likelihood ratios. For one, the \rev{simple-versus-simple} likelihood ratio is an e-variable, and when testing two point hypotheses, it is the unique log-optimal e-variable (see \citet{shafer2021testing} for a simple proof). Second, we can upper bound any e-statistic $E$ for $\calH$ via a likelihood ratio as follows.
Given $P\in\calH$, define a distribution $Q$ via the density $q = p \cdot E / \E_P[E]$ if $\E_P[E]>0$, and $Q = P$ otherwise. Then, since $\E_P[E]\leq 1$,
\begin{equation}\label{eq:late}
    E \leq \frac{q}{p}.
\end{equation}
We have already seen an instance of this in Example~\ref{ex:ryabko}, where the data-compression e-variable was re-expressed as a likelihood ratio.
When $\E_P[E]=1$ (as it is for $E= E_{\tau}$ for all 
\revB{bounded stopping times}
$\tau$ in the e-process in Example~\ref{ex:bounded-means}, for example), \eqref{eq:late} becomes an equality. See \citet[Theorem 14.5]{ramdas2025hypothesis} for a slightly more general statement.

Thus, while e-statistics commit to a distinct core axiom ($\E_P[E]\leq 1$), they often recover the likelihood ratio, or can be expressed or bounded in terms of likelihood ratios.
E-statistics therefore occupy an intermediate position in the debate around Birnbaum's theorem: they are closely connected to likelihood ratios, yet are not defined by a commitment to the likelihood ratio as the canonical measure of evidence. That stance may leave ardent likelihoodists unsatisfied, but others may regard it as a reasonable price to pay for the broader advantages of e-statistics.

\section{Summary and Future Work}

We have positioned e-statistics in the broader literature on statistical evidence. They can be viewed as generalizations of the likelihood ratio, and thus inherit many of the likelihood ratio's appealing evidential properties. E-statistics are fundamentally frequentist objects (though they are allowed to depend on priors via the method of mixtures (Example~\ref{ex:universal-inference}) and recover Bayes factors in particular settings), thus avoiding some of the criticisms of Bayesian confirmation theory.
Their game-theoretic meaning as the wealth accumulated by a statistician betting against nature lends them an intuitive interpretation---one that, like p-values, is well-defined even when considering a single hypothesis. Overall, e-statistics blend several attractive features of p-values, likelihood ratios, and Bayes factors, and we hope to have convinced the reader that they are worthy of consideration as compelling measures of statistical evidence.

With all that said, many evidential aspects of e-statistics remain underexplored. First, \citet{koning2024continuous}, building on work of \citet{ramdas2026randomized}, introduced a novel notion of evidence against $\cH$: the maximum probability with which a certain randomized test, while controlling type-I error, rejects the null. We did not include this notion in our comparison because it differs substantially from existing desiderata, though it may well deserve further study. Second, e-processes as currently defined do not fully satisfy the intra-experiment counterfactual desideratum. Is it possible to generalize their definition so that some do?

\rev{\section*{Acknowledgements}
We thank the anonymous reviewers for helpful comments and J\"urgen Landes for his patience. BC and AR acknowledge support from NSF grants IIS-2229881 and DMS-2310718.}

{\small
\bibliographystyle{plainnat}
\bibliography{main}
}

\end{document}